\documentclass[a4paper,11pt]{article}
\pdfoutput=1
\usepackage{jheppub}

\usepackage{amssymb,amsmath,amsfonts}
\usepackage[normalem]{ulem}
\usepackage[utf8x]{inputenc}
\usepackage{slashed}
\usepackage{graphicx}
\usepackage{tabularx}
\usepackage{here}
\usepackage{color}
\usepackage{csquotes} 
\usepackage{comment}
\usepackage{mathrsfs}
\usepackage{float}
\usepackage{ascmac}
\usepackage{multirow}
\usepackage{longtable}
\usepackage{caption}
\usepackage{bm}
\usepackage{ulem}
\usepackage{tikz}
\usepackage{tikz-feynman}
\tikzfeynmanset{compat=1.1.0, warn luatex=false}
\usetikzlibrary{decorations.pathmorphing,positioning,shapes.geometric,calc,arrows.meta}

\usepackage{colortbl} 

\usepackage[italicdiff]{physics}
\makeatletter
\newcommand*\rel@kern[1]{\kern#1\dimexpr\macc@kerna}
\newcommand*\widebar[1]{%
  \begingroup
  \def\mathaccent##1##2{%
    \rel@kern{0.8}%
    \overline{\rel@kern{-0.8}\macc@nucleus\rel@kern{0.2}}%
    \rel@kern{-0.2}%
  }%
  \macc@depth\@ne
  \let\math@bgroup\@empty \let\math@egroup\macc@set@skewchar
  \mathsurround\z@ \frozen@everymath{\mathgroup\macc@group\relax}%
  \macc@set@skewchar\relax
  \let\mathaccentV\macc@nested@a
  \macc@nested@a\relax111{#1}%
  \endgroup
}
\makeatother

\numberwithin{equation}{section}

\preprint{
\begin{minipage}{5cm}
\small
\flushright
EPHOU-25-009\\
KYUSHU-HET-325
\end{minipage}}

\title{Matter symmetries in supersymmetric standard models from non-invertible
selection rules}

\author{Tatsuo Kobayashi$^{1}$,} 
\author{Hironobu Mita$^{1}$,}
\author{Hajime Otsuka$^{2}$, and}
\author{Riku Sakuma$^{1}$}
\affiliation{
$^1$Department of Physics, Hokkaido University, Sapporo 060-0810, Japan}
\affiliation{
$^2$Department of Physics, Kyushu University, 744 Motooka, Nishi-ku, Fukuoka 819-0395, Japan}
\emailAdd{kobayashi@particle.sci.hokudai.ac.jp}
\emailAdd{h-mita@particle.sci.hokudai.ac.jp}
\emailAdd{otsuka.hajime@phys.kyushu-u.ac.jp}
\emailAdd{r-sakuma@particle.sci.hokudai.ac.jp}

\abstract{
We discuss phenomenological implications of non-invertible selection rules in the framework of the supersymmetric standard model. We find that a remnant $\mathbb{Z}_2$ symmetry of fusion algebras which holds at all-loop order plays the role of $R$-parity, forbidding baryon and lepton violating operators. 
In addition, a combination of Standard Model gauge symmetry and the non-invertible selection rules lead to baryon triality and proton hexality that can protect the proton from decay. In general, our finding selection rules can not realize the same results as conventional $\mathbb{Z}_N$ symmetries in the supersymmetric standard model. 
We also clarify the assignments of matter fields under the fusion algebras that are consistent with 
$SU(4) \times SU(2) \times SU(2)$, $SU(5)$, and $SO(10)$ grand unified theories. 
}

\makeatletter
\gdef\@fpheader{}
\makeatother

\begin{document}

\maketitle

\section{Introduction}

Symmetries are important in physics.
In particle physics, symmetries determine which couplings are allowed or forbidden, that is, the coupling selection rules.
Conventionally, one imposes coupling selection rules due to group theory.
Recently, generalized symmetries have been studied such as non-invertible symmetries.
(See Refs.~\cite{Schafer-Nameki:2023jdn,Shao:2023gho} for reviews of non-invertible symmetries.)
Indeed, non-invertible selection rules appear in various theories. 
For example, they appear in two dimensional conformal field theories~\cite{Bhardwaj:2017xup,Chang:2018iay} and higher dimensional quantum field theories \cite{Heidenreich:2021xpr,Kaidi:2021xfk,Choi:2021kmx}. In the context of string compactifications, non-invertible coupling selection rules appear in heterotic string theory on Calabi-Yau compactifications \cite{Dong:2025pah} and toroidal orbifold compactifications \cite{Dijkgraaf:1987vp,Kobayashi:2004ya, Kobayashi:2006wq,Beye:2014nxa,Thorngren:2021yso,Heckman:2024obe,Kaidi:2024wio}.\footnote{The algebra of vertex operators including twist fields in the worldsheet theory on $S^1/\mathbb{Z}_2$ corresponds to the representation algebra of $D_4$ \cite{Dijkgraaf:1987vp,Beye:2014nxa,Thorngren:2021yso,Kaidi:2024wio,Heckman:2024obe}. Similarly, $\Delta(54)$ symmetry is realized in 
the theory on $T^2/\mathbb{Z}_3$ \cite{Kobayashi:2006wq,Beye:2014nxa}.}
They also appear in type II intersecting/magnetized D-brane models \cite{Kobayashi:2024yqq,Funakoshi:2024uvy}.

Recently, non-invertible symmetries were studied in particle physics \cite{Choi:2022jqy,Cordova:2022fhg,Cordova:2022ieu,Cordova:2024ypu,Kobayashi:2024cvp,Kobayashi:2025znw,Suzuki:2025oov,Liang:2025dkm,Kobayashi:2025ldi,Kobayashi:2025cwx}. 
For example, we can derive phenomenologically interesting textures by using non-invertible selection rules \cite{Kobayashi:2024cvp,Kobayashi:2025znw,Kobayashi:2025ldi}, leading to good agreement between quark/lepton mass matrices and current experimental data. Furthermore, radiative corrections will break the non-invertible selection rules in general. Hence, they play an important role in explaining the origin of tiny neutrino masses, discussed in radiative neutrino mass models \cite{Kobayashi:2025cwx}. It is quite interesting to reveal other phenomenological roles of non-invertible selection rules.

In this paper, we focus on a supersymmetric extension which is one of the promising candidates beyond the standard model.
However, the simple supersymmetric extension of the standard model leads to the fast proton decay problem mediated by supersymmetric partners of the standard model particles.
Usually, some family-independent discrete symmetries are imposed in order to avoid the fast proton decay such as $R$-parity and matter parity ($R_2$) \cite{Farrar:1978xj,Dimopoulos:1981zb,Dimopoulos:1981dw}, 
baryon triality ($B_3$) \cite{Ibanez:1991hv,Ibanez:1991pr}, and 
proton hexality ($P_6$) \cite{Dreiner:2005rd}.

Underlying theory may lead to non-invertible selection rules instead of group-theoretical selection rules.
The purpose of this paper is to study how non-invertible selection rules work as family-independent matter symmetries in the supersymmetric standard model.\footnote{In Refs.~\cite{Kobayashi:2024cvp,Kobayashi:2025znw,Kobayashi:2025ldi}, interesting mass textures were studied by assigning family-dependent classes in $\mathbb{Z}_2$ gauging of $\mathbb{Z}_M$ symmetries. The family-independent matter symmetries, which will be studied in this paper, must be independent of the flavor symmetries studied in Refs.~\cite{Kobayashi:2024cvp,Kobayashi:2025znw,Kobayashi:2025ldi}. We may combine them to realize mass textures and the matter symmetries to avoid the fast proton decay.} 
In particular, we use $\mathbb{Z}_2$ gauging of $\mathbb{Z}_M$ symmetries \cite{Kobayashi:2024yqq,Kobayashi:2024cvp} as non-invertible selection rules including the  Fibonacci fusion rule for $M=3$ and the Ising fusion rule for $M=4$. 
By applying the non-invertible selection rules to the supersymmetric standard model, we find that the effective action enjoys interesting matter symmetries such as the $R$-parity and matter parity ($R_2$) \cite{Farrar:1978xj,Dimopoulos:1981zb,Dimopoulos:1981dw}, 
baryon triality ($B_3$) \cite{Ibanez:1991hv,Ibanez:1991pr}, and 
proton hexality ($P_6$) \cite{Dreiner:2005rd}. Some of matter assignments under the fusion algebras are allowed in $SU(4) \times SU(2) \times SU(2)$, $SU(5)$, and $SO(10)$ grand unified theories. 
Furthermore, there exist $\mathbb{Z}_2$ symmetries originating from the fusion rules, which are expected to be exact at all-loop levels.

This paper is organized as follows.
In section \ref{sec:Z2-gauging}, we briefly review how to derive the coupling selection rules from generic fusion algebras.
In particular, we give a brief review on $\mathbb{Z}_2$ gauging of $\mathbb{Z}_M$ symmetries.
In section \ref{sec:matter-symm}, we review matter $\mathbb{Z}_N$ symmetries in the supersymmetric standard model such as $R_2$, $B_3$, and $P_6$. We classify the remaining matter symmetries in the minimal supersymmetric standard model with the non-invertible selection rules associated with $\mathbb{Z}_2$ gauging of $\mathbb{Z}_3$ and $\mathbb{Z}_4$ symmetries in section~\ref{sec:noninvertible}. 
Section~\ref{sec:con} is devoted to the conclusions. 
In Appendix~\ref{app}, we summarize the remaining matter symmetries in the case of $\mathbb{Z}_2$ gauging of $\mathbb{Z}_5$ symmetry.

\section{$\mathbb{Z}_2$ gauging of $\mathbb{Z}_M$ symmetries}
\label{sec:Z2-gauging}

In group theory, the multiplication law can be written as
\begin{align}
    ab=c,
\end{align}
where $a,b,c$ are elements in the group $G$.
Each field corresponds to a (representation of) group element like $\phi_a, \phi_b, \phi_c$.
The processes $\phi_a + \phi_b \to \phi_c$ is allowed if $ab=c$, but the processes $\phi_a + \phi_b \to \phi_d$ is not allowed if $ab\neq d$.
Another field $\phi'_c$ corresponds to the same $c$ as $\phi_c$, and the processes $\phi_a + \phi_b \to \phi'_c$ is also allowed.

Suppose that an underlying theory leads to the following multiplication rules of elements $U_i$, 
\begin{align}
    U_i U_j = \sum_k c_{ijk} U_k,
\end{align}
where $U_i$ may be (topological) operators.
Then, each field $\phi_i$ corresponds to the element $U_i$.
The processes $\phi_i + \phi_j \to \phi_k$ is allowed if $c_{ijk} \neq 0$, but is forbidden if $c_{ijk} =0$.
Another field $\phi'_k$ can correspond to the same element as $\phi_k$, and the processes are allowed if $c_{ijk} \neq 0$.

Let us discuss such non-trivial multiplication rules realized by $\mathbb{Z}_2$ gauging of $\mathbb{Z}_M$ symmetries \cite{Kobayashi:2024yqq,Kobayashi:2024cvp}. 
We start with the $\mathbb{Z}_M$ symmetry.
We denote its generator by $g$, i.e., $g^M=e$, where $e$ is the identity.
The multiplication law is written as 
\begin{align}
    g^kg^{k'}=g^{k+k'}.
\end{align}

There is the $\mathbb{Z}_2$ outer automorphism,
\begin{align}
    rgr^{-1}=g^{-k},
\end{align}
where $r^2=e$.
By use of this automorphism, we define the class: 
\begin{align}
    [g^k]= \{ hg^kh^{-1} ~|~ h=e, r   \}.
\end{align}
Their multiplication rules can be written by 
\begin{align}\label{eq:fusion-Z2-ZM}
    [g^k][g^{k'}] = [g^{k+k'}] + [g^{M-k+k'}].
\end{align}
That is $\mathbb{Z}_2$ gauging of $\mathbb{Z}_M$ symmetry.

When $M=$ even, the above fusion rule leads to the following law:
\begin{align}\label{eq:fusion-Z2}
    [g^{\rm even}][g^{\rm even}] &= [g^{\rm even}] + [g^{\rm even}],\notag \\
    [g^{\rm even}][g^{\rm odd}] &= [g^{\rm odd}] + [g^{\rm odd}], \\
    [g^{\rm odd}][g^{\rm odd}] &= [g^{\rm even}] + [g^{\rm even}].\notag
\end{align}
Thus, when $M=$ even, the $\mathbb{Z}_2$ symmetry associated with $[g^{\rm odd}]\rightarrow -[g^{\rm odd}]$ appears in the fusion algebra.

$\mathbb{Z}_2$ gauging of $\mathbb{Z}_M$ symmetries can be realized in higher dimensional field theory and string theory.
Certain higher dimensional theories such as magnetized compactification lead to the $\mathbb{Z}_M$ symmetry \cite{Abe:2009vi,Berasaluce-Gonzalez:2012abm,Marchesano:2019hfb}.
The massless modes $\varphi_k$ transform as
\begin{align}
    \varphi_k \to g^k \varphi_k,
\end{align}
where $k$ may correspond to discrete momentum, winding number, or their mixture.
Here, we consider the geometrical $\mathbb{Z}_2$ orbifolding of the above compactification.
Then, the $\mathbb{Z}_2$ invariant modes can be written as \cite{Abe:2008fi}
\begin{align}
    \phi_k = \varphi_k + \varphi_{M-k},
\end{align}
which corresponds to the class $[g^k]$~\cite{Kobayashi:2024yqq,Kobayashi:2024cvp}.
The mode $\phi_k$ behaves as if it has the $k$
 and $M-k$ charges at the same time.

Each field $\phi_k$ in four dimensional effective field theory corresponds to 
a class $[g^k]$.
The two-point coupling of $\phi_{k_1}$ and $\phi_{k_2}$ is allowed if $[g^0]$ appears in the right hand side of the multiplication rule (\ref{eq:fusion-Z2-ZM}) \cite{Kobayashi:2024yqq,Kobayashi:2024cvp}.
Its condition is written by 
\begin{align}
    \pm k_1 \pm k_2=0~~({\rm mod}~~M).
\end{align}
That means that the two-point couplings including mass terms and kinetic terms are allowed between the same class.
Similarly, the n-point coupling $\phi_{k_1} \cdots \phi_{k_n}$ is allowed 
if \cite{Kobayashi:2024yqq,Kobayashi:2024cvp}
\begin{align}
    \sum_{i=1}^n \pm k_i=0~~({\rm mod}~~M).
\end{align}

\section{Matter $\mathbb{Z}_N$ symmetries in supersymmetric models}
\label{sec:matter-symm}

The supersymmetric standard model leads to the fast proton decay.
Several matter $\mathbb{Z}_N$ symmetries have been proposed to avoid the fast proton decay problem.
Among them, famous $\mathbb{Z}_N$ symmetries are $R$-parity and matter parity ($R_2$) \cite{Farrar:1978xj,Dimopoulos:1981zb,Dimopoulos:1981dw}, 
baryon triality ($B_3$) \cite{Ibanez:1991hv,Ibanez:1991pr}, and 
proton hexality ($P_6$) \cite{Dreiner:2005rd}.
Their charge assignments are shown in Table~\ref{tab:matter-ZN}.
We follow the conventional field notations, which would be clear,

All of them allow Yukawa coupling terms in the superpotential:
\begin{align}
\label{eq:YUkawa}
    W=  QH_u \bar U + Q H_d \bar D + L H_d \bar E,
\end{align}
where  
we have omitted flavor indexes and Yukawa coupling constants.
Only the $SU(3) \times SU(2) \times U(1)_Y$ gauge symmetry without the above matter $\mathbb{Z}_N$ symmetries, allow many other terms including dimension 5 operators, which are shown in the first rows of Tables~\ref{tab:allowd-op} and \ref{tab:allowd-op-D}. 
In addition, Tables~\ref{tab:allowd-op} and \ref{tab:allowd-op-D} show whether these operators are allowed or not by the $\mathbb{Z}_N$ symmetries.
The check symbol $\checkmark$ denotes the allowed operator by the $\mathbb{Z}_N$ symmetry.
All of three $\mathbb{Z}_N$ symmetries allow the $\mu$-term $H_uH_d$ and the Weinberg operator $LH_uLH_u$.
The other operators are forbidden by the proton hexality $P_6$.
All of the dimension 4 operators except Yukawa terms are forbidden by $R_2$, but some of the dimension 5 operators are allowed.

\begin{table}[H]
  \centering
  \caption{Matter $\mathbb{Z}_N$ symmetries.}
  \label{tab:matter-ZN}
  \begin{tabular}{|l |c |c |c |c |c |c |c|}
    \hline
    Field & $Q$ & $H_u$ & $H_d$ & $\bar{U}$ & $\bar{D}$ & $\bar{E}$ & $L$ \\
    \hline
    \rowcolor{red!20}
    $R_2$& 1 & 0 & 0 & 1 & 1 & 1 & 1 \\
    \hline
    \rowcolor{blue!20}
    $B_3$ & 0 & 1 & -1 & -1 & 1 & 2 & -1 \\
    \hline
    \rowcolor{green!20}
    $P_6$& 0 & -1 & 1 & 1 & -1 & 1 & -2 \\
    \hline
  \end{tabular}
\end{table}

\begin{table}[H]
  \centering
  \caption{Allowed operators from F-terms by matter $\mathbb{Z}_N$ symmetries. The check symbol $\checkmark$ denotes the allowed operators.}
  \label{tab:allowd-op}
  \footnotesize
  \begin{tabular}{|l  |c |c |c |c |c ||c |c |c |c |c |c|}
    \hline
     &  \hspace{-.1cm} $H_dH_u$ \hspace{-.1cm} & \hspace{-.1cm}$LH_u$\hspace{-.1cm} &\hspace{-.1cm} $LL\bar{E}$ \hspace{-.1cm} &\hspace{-.1cm} $LQ\bar{D}$ \hspace{-.1cm} & \hspace{-.1cm}$\bar{U}\bar{D}\bar{D}$\hspace{-.1cm} & \hspace{-.1cm}$QQQL$\hspace{-.1cm} & \hspace{-.1cm}$\bar{U}\bar{U}\bar{D}\bar{E}$\hspace{-.1cm} & \hspace{-.1cm}$QQQH_d$\hspace{-.1cm} & \hspace{-.1cm}$Q\bar{U}\bar{E}H_d$\hspace{-.1cm} & \hspace{-.1cm}$LH_uLH_u$ & $LH_uH_dH_u$\hspace{-.1cm} \\
    \hline
    \rowcolor{red!20}
    $R_2$  & \checkmark &  &  &  &  &\checkmark & \checkmark &  &  & \checkmark &  \\
    \hline
    \rowcolor{blue!20}
    $B_3$  & \checkmark & \checkmark & \checkmark & \checkmark &  &  &  &  & \checkmark & \checkmark & \checkmark \\
    \hline
    \rowcolor{green!20}
    $P_6$  & \checkmark &  & &  &  &  &  &  &  & \checkmark & \\
    \hline
  \end{tabular}
\end{table}

\begin{table}[H]
  \centering
  \caption{Allowed operators from D-terms by matter $\mathbb{Z}_N$ symmetries. The check symbol $\checkmark$ denotes the allowed operators.}
  \label{tab:allowd-op-D}
  \footnotesize
  \begin{tabular}{|l  |c |c |c |c |}
    \hline
     &  \hspace{-.1cm} $\bar U \bar D^* \bar E$ \hspace{-.1cm} & \hspace{-.1cm}$H^*_u H_d \bar E$\hspace{-.1cm} &\hspace{-.1cm} $Q \bar U L^*$ \hspace{-.1cm} &\hspace{-.1cm} $QQ\bar D^*$ \hspace{-.1cm}  \\
    \hline
    \rowcolor{red!20}
    $R_2$  &  &  & & \\
    \hline
    \rowcolor{blue!20}
    $B_3$  & \checkmark & \checkmark & \checkmark &  \\
    \hline
    \rowcolor{green!20}
    $P_6$    & &  & &  \\
    \hline
  \end{tabular}
\end{table}

\section{Non-invertible selection rules}
\label{sec:noninvertible}

In this section, we apply coupling selection rules by $\mathbb{Z}_2$ gauging of the $\mathbb{Z}_M$ symmetries to the supersymmetric standard model. 
Let us start with $M=3$. 
There are two classes, $[g^0]$ and $[g^1]$ obeying the following fusion rule:
\begin{align}
    [g^0] [g^0] &= [g^0]\,,
\quad
    [g^0] [g^1] = [g^1] [g^0] = [g^1]\,,
\quad
    [g^1] [g^1] = [g^0] + [g^1]\,,
\end{align}
i.e., the so-called Fibonacci fusion rule. 
We assign these two classes to chiral matter fields, $Q$, $\bar U$, $\bar D$, $L$, $\bar E$, $H_u$, and $H_d$.
We restrict ourselves to the assignments to allow Yukawa couplings (\ref{eq:YUkawa}).
All possible assignments are shown in Table \ref{tab:assign-M=3}.
Tables \ref{tab:allowed-op-M=3} and \ref{tab:allow-op-M=3D} respectively show allowed couplings in F-term and D-term by each assignment.

\begin{table}[H]
    \centering
    \caption{Possible assignments to allow Yukawa couplings terms for $M=3$, where the symbol $\checkmark$ denotes the consistent assignments of matter fields under $SU(5)$, $SU(4) \times SU(2) \times SU(2)$, and $SO(10)$ grand unified theories, represented by SU5, PS, and SO10 respectively.}
    \label{tab:assign-M=3}
    \begin{longtable}{|l |c |c |c |c |c |c |c |c |c |c |}
        \hline
        Case & $Q$ & $H_u$ & $H_d$ & $\bar{U}$ & $\bar{D}$ & $\bar{E}$ & $L$ & SU5 & PS & SO10 \\
        \hline
        \endfirsthead
        \hline
        Case & $Q$ & $H_u$ & $H_d$ & $\bar{U}$ & $\bar{D}$ & $\bar{E}$ & $L$ & SU5 & PS & SO10 \\
        \hline
        \endhead
        ${M=3}$ (1) & 0 & 0 & 0 & 0 & 0 & 0 & 0 & \checkmark & \checkmark & \checkmark \\
        \hline
        ${M=3}$ (2) & 0 & 0 & 0 & 0 & 0 & 1 & 1 &  &  &  \\
        \hline
        ${M=3}$ (3) & 0 & 0 & 1 & 0 & 1 & 0 & 1 & \checkmark &  &  \\
        \hline
        ${M=3}$ (4) & 0 & 0 & 1 & 0 & 1 & 1 & 0 &  &  &  \\
        \hline
        ${M=3}$ (5) & 0 & 0 & 1 & 0 & 1 & 1 & 1 &  &  &  \\
        \hline
        ${M=3}$ (6) & 0 & 1 & 0 & 1 & 0 & 0 & 0 &  &  &  \\
        \hline
        ${M=3}$ (7) & 0 & 1 & 0 & 1 & 0 & 1 & 1 &  &  &  \\
        \hline
        ${M=3}$ (8) & 0 & 1 & 1 & 1 & 1 & 0 & 1 &  &  &  \\
        \hline
        ${M=3}$ (9) & 0 & 1 & 1 & 1 & 1 & 1 & 0 &  & \checkmark &  \\
        \hline
        ${M=3}$ (10) & 0 & 1 & 1 & 1 & 1 & 1 & 1 &  &  &  \\
        \hline
        ${M=3}$ (11) & 1 & 0 & 0 & 1 & 1 & 0 & 0 &  &  &  \\
        \hline
        ${M=3}$ (12) & 1 & 0 & 0 & 1 & 1 & 1 & 1 & \checkmark & \checkmark & \checkmark \\
        \hline
        ${M=3}$ (13) & 1 & 0 & 1 & 1 & 0 & 0 & 1 &  &  &  \\
        \hline
        ${M=3}$ (14) & 1 & 0 & 1 & 1 & 0 & 1 & 0 & \checkmark &  &  \\
        \hline
        ${M=3}$ (15) & 1 & 0 & 1 & 1 & 0 & 1 & 1 &  &  &  \\
        \hline
        ${M=3}$ (16) & 1 & 0 & 1 & 1 & 1 & 0 & 1 &  &  &  \\
        \hline
        ${M=3}$ (17) & 1 & 0 & 1 & 1 & 1 & 1 & 0 &  &  &  \\
        \hline
        ${M=3}$ (18) & 1 & 0 & 1 & 1 & 1 & 1 & 1 & \checkmark & \checkmark &  \\
        \hline
        ${M=3}$ (19) & 1 & 1 & 0 & 0 & 1 & 0 & 0 &  &  &  \\
        \hline
        ${M=3}$ (20) & 1 & 1 & 0 & 0 & 1 & 1 & 1 &  &  &  \\
        \hline
        ${M=3}$ (21) & 1 & 1 & 0 & 1 & 1 & 0 & 0 &  &  &  \\
        \hline
        ${M=3}$ (22) & 1 & 1 & 0 & 1 & 1 & 1 & 1 & \checkmark & \checkmark &  \\
        \hline
        ${M=3}$ (23) & 1 & 1 & 1 & 0 & 0 & 0 & 1 &  & \checkmark &  \\
        \hline
        ${M=3}$ (24) & 1 & 1 & 1 & 0 & 0 & 1 & 0 &  &  &  \\
        \hline
        ${M=3}$ (25) & 1 & 1 & 1 & 0 & 0 & 1 & 1 &  &  &  \\
        \hline
        ${M=3}$ (26) & 1 & 1 & 1 & 0 & 1 & 0 & 1 &  &  &  \\
        \hline
        ${M=3}$ (27) & 1 & 1 & 1 & 0 & 1 & 1 & 0 &  &  &  \\
        \hline
        ${M=3}$ (28) & 1 & 1 & 1 & 0 & 1 & 1 & 1 &  &  &  \\
        \hline
        ${M=3}$ (29) & 1 & 1 & 1 & 1 & 0 & 0 & 1 &  &  &  \\
        \hline
        ${M=3}$ (30) & 1 & 1 & 1 & 1 & 0 & 1 & 0 & \checkmark &  &  \\
        \hline
        ${M=3}$ (31) & 1 & 1 & 1 & 1 & 0 & 1 & 1 &  &  &  \\
        \hline
        ${M=3}$ (32) & 1 & 1 & 1 & 1 & 1 & 0 & 1 &  &  &  \\
        \hline
        ${M=3}$ (33) & 1 & 1 & 1 & 1 & 1 & 1 & 0 &  &  &  \\
        \hline
        ${M=3}$ (34) & 1 & 1 & 1 & 1 & 1 & 1 & 1 & \checkmark & \checkmark & \checkmark \\
        \hline
    \end{longtable}
\end{table}

\setcounter{table}{4}

\begin{table}[H]
    \centering
    \caption{Allowed operators in F-terms for $M=3$, represented by the symbol $\checkmark$ for each case in Table~\ref{tab:assign-M=3}.}
    \label{tab:allowed-op-M=3}
    \scriptsize
    \begin{longtable}{|p{0.6cm} |c |c |c |c |c ||c |c |c |c |c|}
    \hline
        Case & $H_dH_u$ & $LH_u$ & $LL\bar{E}$ & $LQ\bar{D}$ & $\bar{U}\bar{D}\bar{D}$ & $QQQL$ & $\bar{U}\bar{U}\bar{D}\bar{E}$ & $QQQH_d$ & $Q\bar{U}\bar{E}H_d$  & $LH_uH_dH_u$ \\
        \hline\hline
        \endfirsthead
        
        \hline
        Case & $H_dH_u$ & $LH_u$ & $LL\bar{E}$ & $LQ\bar{D}$ & $\bar{U}\bar{D}\bar{D}$ & $QQQL$ & $\bar{U}\bar{U}\bar{D}\bar{E}$ & $QQQH_d$ & $Q\bar{U}\bar{E}H_d$ & $LH_uH_dH_u$ \\
        \hline\hline
        \endhead
        (1), (11), (23), 
        (28), (32), (34)
        & \checkmark & \checkmark & \checkmark & \checkmark & \checkmark & \checkmark & \checkmark & \checkmark & \checkmark  & \checkmark \\
        \hline
        (2) & \checkmark &  & \checkmark &  & \checkmark &  &  & \checkmark &  &  \\
        \hline
        (3) &  &  & \checkmark & \checkmark & \checkmark &  &  &  &  & \checkmark \\
        \hline
        (4) &  & \checkmark &  &  & \checkmark & \checkmark & \checkmark &  & \checkmark  &  \\
        \hline
        (5) &  &  & \checkmark & \checkmark & \checkmark &  & \checkmark &  & \checkmark &  \checkmark \\
        \hline
        (6) &  &  & \checkmark & \checkmark &  & \checkmark & \checkmark & \checkmark &   & \checkmark \\
        \hline
        (7) &  & \checkmark & \checkmark &  &  &  & \checkmark & \checkmark & \checkmark & \checkmark \\
        \hline
        (8), (10) & \checkmark & \checkmark & \checkmark & \checkmark & \checkmark &  & \checkmark &  & \checkmark & \checkmark \\
        \hline
        (9) & \checkmark &  &  &  & \checkmark & \checkmark & \checkmark &  & \checkmark & \checkmark \\
        \hline
        (12) & \checkmark &  & \checkmark & \checkmark & \checkmark & \checkmark & \checkmark & \checkmark & \checkmark &  \\
        \hline
        (13), (15) &  &  & \checkmark & \checkmark &  & \checkmark & \checkmark & \checkmark & \checkmark & \checkmark \\
        \hline
        (14) &  & \checkmark &  &  &  & \checkmark & \checkmark & \checkmark & \checkmark &  \\
        \hline
        (16), (18), (21) 
        &  &  & \checkmark & \checkmark & \checkmark & \checkmark & \checkmark & \checkmark & \checkmark  & \checkmark \\
        \hline
        (17) &  & \checkmark &  & \checkmark & \checkmark & \checkmark & \checkmark & \checkmark & \checkmark &   \\
        \hline
        (19) &  &  & \checkmark & \checkmark & \checkmark & \checkmark &  & \checkmark &  & \checkmark \\
        \hline
        (20), (22) &  & \checkmark & \checkmark & \checkmark & \checkmark & \checkmark & \checkmark & \checkmark & \checkmark  & \checkmark \\
        \hline
        (24) & \checkmark &  &  &  & \checkmark & \checkmark &  & \checkmark & \checkmark  & \checkmark \\
        \hline
        (25), (26) & \checkmark & \checkmark & \checkmark & \checkmark & \checkmark & \checkmark &  & \checkmark & \checkmark & \checkmark \\
        \hline
        (27), (33) & \checkmark &  &  & \checkmark & \checkmark & \checkmark & \checkmark & \checkmark & \checkmark & \checkmark \\
        \hline
        (29), (31) & \checkmark & \checkmark & \checkmark & \checkmark &  & \checkmark & \checkmark & \checkmark & \checkmark & \checkmark \\
        \hline
        (30) & \checkmark &  &  &  &  & \checkmark & \checkmark & \checkmark & \checkmark & \checkmark \\
        \hline
    \end{longtable}  
\end{table}

\setcounter{table}{5}

\begin{table}[H]
    \centering
  \caption{Allowed operators in D-terms represented by the symbol $\checkmark$ for $M=3$.}
    \label{tab:allow-op-M=3D}
    \begin{longtable}{|p{2.5cm} |c |c |c |c|}
        \hline
        Case & $\bar{U}\bar{D}^*\bar{E}$ & $H_u^*H_d\bar{E}$ & $Q\bar{U}L^*$ & $QQ\bar{D}^*$ \\\hline
        \endfirsthead
        \hline
        Case & $\bar{U}\bar{D}^*\bar{E}$ & $H_u^*H_d\bar{E}$ & $Q\bar{U}L^*$ & $QQ\bar{D}^*$ \\\hline
        \endhead
        \hline
        \endfoot
        \hline
        \endlastfoot
        (1), (7), (11), (14), (15), (17), (18), (20), (22), (23), (28), (30), (31), (32), (33), (34) & \checkmark & \checkmark & \checkmark & \checkmark \\\hline
        (2), (6), (19) &  &  &  & \checkmark \\\hline
        (3) &  &  &  &  \\\hline
        (4), (8), (10) & \checkmark & \checkmark & \checkmark &  \\\hline
        (5), (9) & \checkmark & \checkmark &  &  \\\hline
        (12), (16), (21) & \checkmark &  & \checkmark & \checkmark \\\hline
        (13) &  &  & \checkmark & \checkmark \\\hline
        (24) &  & \checkmark &  & \checkmark \\\hline
        (25), (26), (29) &  & \checkmark & \checkmark & \checkmark \\\hline
        (27) & \checkmark & \checkmark &  & \checkmark \\\hline
    \end{longtable}
\end{table}

Similarly, we can study the coupling selection rules due to $\mathbb{Z}_2$ gauging of $\mathbb{Z}_4$ symmetry.
There are three classes, $[g^0]$, $[g^1]$, and $[g^2]$, obeying the following fusion rule:
\begin{align}
\label{eq:Ising}
    [g^0] [g^0] &= [g^0]\,,
    \qquad\qquad
    [g^0] [g^i] = [g^i] [g^0] = [g^i]\,,
    \nonumber\\
    [g^1] [g^1] &= [g^0] + [g^2]\,,
    \quad \,
    [g^1] [g^2] = [g^2] [g^1] = [g^1]\,,\\
    [g^2] [g^2] &= [g^0]\,,\nonumber
\end{align}
with $i=1,2$, i.e., the so-called Ising fusion rule. 
Table \ref{tab:assign-M=4} shows all the possible assignments of classes, which allow Yukawa couplings (\ref{eq:YUkawa}). 
Tables \ref{tab:allowed-op-M=4} and \ref{tab:allowed-op-M=4D} respectively show allowed couplings in F-term and D-term by each assignment.

\newpage

\setcounter{table}{6}

\begin{longtable}{|l |c |c |c |c |c |c |c |c |c |c |c |}
    \caption{Possible assignments to allow Yukawa couplings for $M=4$, where the symbol $\checkmark$ denotes the consistent assignments of matter fields under $SU(5)$, $SU(4) \times SU(2) \times SU(2)$, and $SO(10)$ grand unified theories, represented by SU5, PS, and SO10 respectively. In the last column, the symbol $\checkmark$ or $\ast$ is introduced if the $\mathbb{Z}_2$ symmetry exists.}
    \label{tab:assign-M=4} 
\\
    \hline
    Case & $Q$ & $H_u$ & $H_d$ & $\bar{U}$ & $\bar{D}$ & $\bar{E}$ & $L$ & SU5 & PS & SO10 & $\mathbb{Z}_2$ \\
    \hline
    \endfirsthead
    \hline
    Case & $Q$ & $H_u$ & $H_d$ & $\bar{U}$ & $\bar{D}$ & $\bar{E}$ & $L$ & SU5 & PS & SO10 & $\mathbb{Z}_2$ \\
    \hline
    \endhead
    ${M=4}$ (1) & 0 & 0 & 0 & 0 & 0 & 0 & 0 & \checkmark & \checkmark & \checkmark& $*$ \\
    \hline
    ${M=4}$ (2) & 0 & 0 & 0 & 0 & 0 & 1 & 1 &  &  & &\checkmark \\
    \hline
    ${M=4}$ (3) & 0 & 0 & 0 & 0 & 0 & 2 & 2 &  &  & & $*$\\
    \hline
    ${M=4}$ (4) & 0 & 0 & 1 & 0 & 1 & 0 & 1 & \checkmark &  & & \checkmark\\
    \hline
    ${M=4}$ (5) & 0 & 0 & 1 & 0 & 1 & 1 & 0 &  &  & & \checkmark\\
    \hline
    ${M=4}$ (6) & 0 & 0 & 1 & 0 & 1 & 1 & 2 &  &  & & \\
    \hline
    ${M=4}$ (7) & 0 & 0 & 1 & 0 & 1 & 2 & 1 &  &  & & \checkmark\\
    \hline
    ${M=4}$ (8) & 0 & 0 & 2 & 0 & 2 & 0 & 2 & \checkmark &  & & $*$\\
    \hline
    ${M=4}$ (9) & 0 & 0 & 2 & 0 & 2 & 1 & 1 &  &  & & \\
    \hline
    ${M=4}$ (10) & 0 & 0 & 2 & 0 & 2 & 2 & 0 &  &  & & $*$\\
    \hline
    ${M=4}$ (11) & 0 & 1 & 0 & 1 & 0 & 0 & 0 &  &  & & \checkmark\\
    \hline
    ${M=4}$ (12) & 0 & 1 & 0 & 1 & 0 & 1 & 1 &  &  &  & \checkmark\\
    \hline
    ${M=4}$ (13) & 0 & 1 & 0 & 1 & 0 & 2 & 2 &  &  & & \\
    \hline
    \rowcolor{blue!20}${M=4}$ (14) & 0 & 1 & 1 & 1 & 1 & 0 & 1 &  &  & &\checkmark \\
    \hline
    \rowcolor{red!20}${M=4}$ (15) & 0 & 1 & 1 & 1 & 1 & 1 & 0 &  & \checkmark &  &\checkmark \\
    \hline
    ${M=4}$ (16) & 0 & 1 & 1 & 1 & 1 & 1 & 2 &  &  & & \\
    \hline
    \rowcolor{blue!20}${M=4}$ (17) & 0 & 1 & 1 & 1 & 1 & 2 & 1 &  &  & & \checkmark\\
    \hline
    ${M=4}$ (18) & 0 & 1 & 2 & 1 & 2 & 0 & 2 &  &  & & \\
    \hline
    ${M=4}$ (19) & 0 & 1 & 2 & 1 & 2 & 1 & 1 &  &  & & \\
    \hline
    ${M=4}$ (20) & 0 & 1 & 2 & 1 & 2 & 2 & 0 &  &  & & \\
    \hline
    ${M=4}$ (21) & 0 & 2 & 0 & 2 & 0 & 0 & 0 &  &  &  & $*$\\
    \hline
    ${M=4}$ (22) & 0 & 2 & 0 & 2 & 0 & 1 & 1 &  &  & & \\
    \hline
    ${M=4}$ (23) & 0 & 2 & 0 & 2 & 0 & 2 & 2 &  &  &  & $*$\\
    \hline
    ${M=4}$ (24) & 0 & 2 & 1 & 2 & 1 & 0 & 1 &  &  &  & \checkmark\\
    \hline
    ${M=4}$ (25) & 0 & 2 & 1 & 2 & 1 & 1 & 0 &  &  & & \\
    \hline
    ${M=4}$ (26) & 0 & 2 & 1 & 2 & 1 & 1 & 2 &  &  & & \\
    \hline
    ${M=4}$ (27) & 0 & 2 & 1 & 2 & 1 & 2 & 1 &  &  &  & \checkmark\\
    \hline
    \rowcolor{blue!20}${M=4}$ (28) & 0 & 2 & 2 & 2 & 2 & 0 & 2 &  &  &  & $*$\\
    \hline
    \rowcolor{green!20}${M=4}$ (29) & 0 & 2 & 2 & 2 & 2 & 1 & 1 &  &  & & \\
    \hline
    \rowcolor{red!20}${M=4}$ (30) & 0 & 2 & 2 & 2 & 2 & 2 & 0 &  & \checkmark &  & $*$\\
    \hline
    \rowcolor{blue!20}${M=4}$ (31) & 1 & 0 & 0 & 1 & 1 & 0 & 0 &  &  &  &\checkmark \\
    \hline
    \rowcolor{red!20}${M=4}$ (32) & 1 & 0 & 0 & 1 & 1 & 1 & 1 & \checkmark & \checkmark & \checkmark &\checkmark \\
    \hline
    ${M=4}$ (33) & 1 & 0 & 0 & 1 & 1 & 2 & 2 &  &  &  &\\
    \hline
    ${M=4}$ (34) & 1 & 0 & 1 & 1 & 0 & 0 & 1 &  &  &  & \checkmark\\
    \hline
    ${M=4}$ (35) & 1 & 0 & 1 & 1 & 0 & 1 & 0 & \checkmark &  &  & \checkmark\\
    \hline
    ${M=4}$ (36) & 1 & 0 & 1 & 1 & 0 & 1 & 2 &  &  &  &\\
    \hline
    ${M=4}$ (37) & 1 & 0 & 1 & 1 & 0 & 2 & 1 &  &  &  & \checkmark\\
    \hline
    ${M=4}$ (38) & 1 & 0 & 1 & 1 & 2 & 0 & 1 &  &  &  & \checkmark\\
    \hline
    ${M=4}$ (39) & 1 & 0 & 1 & 1 & 2 & 1 & 0 &  &  &  & \checkmark\\
    \hline
    ${M=4}$ (40) & 1 & 0 & 1 & 1 & 2 & 1 & 2 & \checkmark &  &  &\\
    \hline
    ${M=4}$ (41) & 1 & 0 & 1 & 1 & 2 & 2 & 1 &  &  &  & \checkmark\\
    \hline
    ${M=4}$ (42) & 1 & 0 & 2 & 1 & 1 & 0 & 2 &  &  &  &\\
    \hline
    \rowcolor{red!20}${M=4}$ (43) & 1 & 0 & 2 & 1 & 1 & 1 & 1 & \checkmark & \checkmark &  &\\
    \hline
    ${M=4}$ (44) & 1 & 0 & 2 & 1 & 1 & 2 & 0 &  &  &  &\\
    \hline
    ${M=4}$ (45) & 1 & 1 & 0 & 0 & 1 & 0 & 0 &  &  &  & \checkmark\\
    \hline
    ${M=4}$ (46) & 1 & 1 & 0 & 0 & 1 & 1 & 1 &  &  &  & \checkmark\\
    \hline
    ${M=4}$ (47) & 1 & 1 & 0 & 0 & 1 & 2 & 2 &  &  &  &\\
    \hline
    ${M=4}$ (48) & 1 & 1 & 0 & 2 & 1 & 0 & 0 &  &  & &  \checkmark\\
    \hline
    ${M=4}$ (49) & 1 & 1 & 0 & 2 & 1 & 1 & 1 &  &  &  & \checkmark\\
    \hline
    ${M=4}$ (50) & 1 & 1 & 0 & 2 & 1 & 2 & 2 &  &  &  &\\
    \hline
    ${M=4}$ (51) & 1 & 1 & 1 & 0 & 0 & 0 & 1 &  & \checkmark &  & \checkmark\\
    \hline
    ${M=4}$ (52) & 1 & 1 & 1 & 0 & 0 & 1 & 0 &  &  &  & \checkmark\\
    \hline
    ${M=4}$ (53) & 1 & 1 & 1 & 0 & 0 & 1 & 2 &  &  &  & \checkmark\\
    \hline
    ${M=4}$ (54) & 1 & 1 & 1 & 0 & 0 & 2 & 1 &  &  &  &\\
    \hline
    ${M=4}$ (55) & 1 & 1 & 1 & 0 & 2 & 0 & 1 &  &  &  &\\
    \hline
    ${M=4}$ (56) & 1 & 1 & 1 & 0 & 2 & 1 & 0 &  &  &  & \checkmark\\
    \hline
    ${M=4}$ (57) & 1 & 1 & 1 & 0 & 2 & 1 & 2 &  &  &  & \checkmark\\
    \hline
    ${M=4}$ (58) & 1 & 1 & 1 & 0 & 2 & 2 & 1 &  &  &  & \checkmark\\
    \hline
    ${M=4}$ (59) & 1 & 1 & 1 & 2 & 0 & 0 & 1 &  &  &  &\\
    \hline
    ${M=4}$ (60) & 1 & 1 & 1 & 2 & 0 & 1 & 0 &  &  &  &\\
    \hline
    ${M=4}$ (61) & 1 & 1 & 1 & 2 & 0 & 1 & 2 &  &  &  &\\
    \hline
    ${M=4}$ (62) & 1 & 1 & 1 & 2 & 0 & 2 & 1 &  &  &  &\\
    \hline
    ${M=4}$ (63) & 1 & 1 & 1 & 2 & 2 & 0 & 1 &  &  &  &\\
    \hline
    ${M=4}$ (64) & 1 & 1 & 1 & 2 & 2 & 1 & 0 &  &  & & \\
    \hline
    ${M=4}$ (65) & 1 & 1 & 1 & 2 & 2 & 1 & 2 &  &  &  &\\
    \hline
    ${M=4}$ (66) & 1 & 1 & 1 & 2 & 2 & 2 & 1 &  & \checkmark &  &\\
    \hline
    ${M=4}$ (67) & 1 & 1 & 2 & 0 & 1 & 0 & 2 &  &  & & \checkmark\\
    \hline
    ${M=4}$ (68) & 1 & 1 & 2 & 0 & 1 & 1 & 1 &  &  &  & \checkmark\\
    \hline
    ${M=4}$ (69) & 1 & 1 & 2 & 0 & 1 & 2 & 0 &  &  & & \\
    \hline
    ${M=4}$ (70) & 1 & 1 & 2 & 2 & 1 & 0 & 2 &  &  & & \checkmark\\
    \hline
    ${M=4}$ (71) & 1 & 1 & 2 & 2 & 1 & 1 & 1 &  &  & & \checkmark\\
    \hline
    ${M=4}$ (72) & 1 & 1 & 2 & 2 & 1 & 2 & 0 &  &  & & \\
    \hline
    ${M=4}$ (73) & 1 & 2 & 0 & 1 & 1 & 0 & 0 &  &  &  &\\
    \hline
    \rowcolor{red!20}${M=4}$ (74) & 1 & 2 & 0 & 1 & 1 & 1 & 1 & \checkmark & \checkmark & & \\
    \hline
    ${M=4}$ (75) & 1 & 2 & 0 & 1 & 1 & 2 & 2 &  &  & & \\
    \hline
    ${M=4}$ (76) & 1 & 2 & 1 & 1 & 0 & 0 & 1 &  &  & & \checkmark\\
    \hline
    ${M=4}$ (77) & 1 & 2 & 1 & 1 & 0 & 1 & 0 & \checkmark &  &  &\\
    \hline
    ${M=4}$ (78) & 1 & 2 & 1 & 1 & 0 & 1 & 2 &  &  & & \checkmark\\
    \hline
    ${M=4}$ (79) & 1 & 2 & 1 & 1 & 0 & 2 & 1 &  &  & & \checkmark\\
    \hline
    ${M=4}$ (80) & 1 & 2 & 1 & 1 & 2 & 0 & 1 &  &  & & \checkmark\\
    \hline
    ${M=4}$ (81) & 1 & 2 & 1 & 1 & 2 & 1 & 0 &  &  & & \\
    \hline
    ${M=4}$ (82) & 1 & 2 & 1 & 1 & 2 & 1 & 2 & \checkmark &  &  & \checkmark\\
    \hline
    ${M=4}$ (83) & 1 & 2 & 1 & 1 & 2 & 2 & 1 &  &  &  & \checkmark\\
    \hline
    \rowcolor{blue!20}${M=4}$ (84) & 1 & 2 & 2 & 1 & 1 & 0 & 2 &  &  &  & \checkmark \\
    \hline
    \rowcolor{red!20}${M=4}$ (85) & 1 & 2 & 2 & 1 & 1 & 1 & 1 & \checkmark & \checkmark & \checkmark &\checkmark \\
    \hline
    ${M=4}$ (86) & 1 & 2 & 2 & 1 & 1 & 2 & 0 &  &  &  &\\
    \hline
    \rowcolor{blue!20}${M=4}$ (87) & 2 & 0 & 0 & 2 & 2 & 0 & 0 &  &  & & $*$\\
    \hline
    \rowcolor{green!20}${M=4}$ (88) & 2 & 0 & 0 & 2 & 2 & 1 & 1 &  &  &  &\\
    \hline
    \rowcolor{red!20}${M=4}$ (89) & 2 & 0 & 0 & 2 & 2 & 2 & 2 & \checkmark & \checkmark & \checkmark &\\
    \hline
    ${M=4}$ (90) & 2 & 0 & 1 & 2 & 1 & 0 & 1 &  &  &  & \checkmark\\
    \hline
    ${M=4}$ (91) & 2 & 0 & 1 & 2 & 1 & 1 & 0 &  &  &  &\\
    \hline
    ${M=4}$ (92) & 2 & 0 & 1 & 2 & 1 & 1 & 2 &  &  &  &\\
    \hline
    ${M=4}$ (93) & 2 & 0 & 1 & 2 & 1 & 2 & 1 & \checkmark &  &  & \checkmark\\
    \hline
    ${M=4}$ (94) & 2 & 0 & 2 & 2 & 0 & 0 & 2 &  &  &  &\\
    \hline
    ${M=4}$ (95) & 2 & 0 & 2 & 2 & 0 & 1 & 1 &  &  & & \\
    \hline
    ${M=4}$ (96) & 2 & 0 & 2 & 2 & 0 & 2 & 0 & \checkmark &  &  & $*$\\
    \hline
    ${M=4}$ (97) & 2 & 1 & 0 & 1 & 2 & 0 & 0 &  &  & & \\
    \hline
    ${M=4}$ (98) & 2 & 1 & 0 & 1 & 2 & 1 & 1 &  &  &  &\\
    \hline
    ${M=4}$ (99) & 2 & 1 & 0 & 1 & 2 & 2 & 2 &  &  &  &\\
    \hline
    \rowcolor{blue!20}${M=4}$ (100) & 2 & 1 & 1 & 1 & 1 & 0 & 1 &  &  &  &\checkmark \\
    \hline
    ${M=4}$ (101) & 2 & 1 & 1 & 1 & 1 & 1 & 0 &  &  &  &\\
    \hline
    \rowcolor{red!20}${M=4}$ (102) & 2 & 1 & 1 & 1 & 1 & 1 & 2 &  & \checkmark &  &\checkmark \\
    \hline
    \rowcolor{blue!20}${M=4}$ (103) & 2 & 1 & 1 & 1 & 1 & 2 & 1 &  &  &  &\checkmark \\
    \hline
    ${M=4}$ (104) & 2 & 1 & 2 & 1 & 0 & 0 & 2 &  &  &  & \checkmark\\
    \hline
    ${M=4}$ (105) & 2 & 1 & 2 & 1 & 0 & 1 & 1 &  &  &  & \checkmark\\
    \hline
    ${M=4}$ (106) & 2 & 1 & 2 & 1 & 0 & 2 & 0 &  &  &  &\\
    \hline
    ${M=4}$ (107) & 2 & 2 & 0 & 0 & 2 & 0 & 0 &  &  &  &\\
    \hline
    ${M=4}$ (108) & 2 & 2 & 0 & 0 & 2 & 1 & 1 &  &  &  &\\
    \hline
    ${M=4}$ (109) & 2 & 2 & 0 & 0 & 2 & 2 & 2 &  &  &  &\\
    \hline
    ${M=4}$ (110) & 2 & 2 & 1 & 0 & 1 & 0 & 1 &  &  &  & \checkmark\\
    \hline
    ${M=4}$ (111) & 2 & 2 & 1 & 0 & 1 & 1 & 0 &  &  &  &\\
    \hline
    ${M=4}$ (112) & 2 & 2 & 1 & 0 & 1 & 1 & 2 &  &  &  & \checkmark\\
    \hline
    ${M=4}$ (113) & 2 & 2 & 1 & 0 & 1 & 2 & 1 &  &  &  & \checkmark\\
    \hline
    ${M=4}$ (114) & 2 & 2 & 2 & 0 & 0 & 0 & 2 &  & \checkmark &  & $*$ \\
    \hline
    ${M=4}$ (115) & 2 & 2 & 2 & 0 & 0 & 1 & 1 &  &  &  & \checkmark\\
    \hline
    ${M=4}$ (116) & 2 & 2 & 2 & 0 & 0 & 2 & 0 &  &  &  & $*$\\
    \hline
    \end{longtable}
\newpage

\begin{tiny}
\begin{longtable}{|p{1.5cm} |c |c |c |c |c ||c |c |c |c |c |c|}
    \caption{Allowed operators for $M=4$, represented by the symbol $\checkmark$ for each case in Table~\ref{tab:assign-M=4}.}
    \label{tab:allowed-op-M=4}  \\
    \hline
    Case & $H_dH_u$ & $LH_u$ & $LL\bar{E}$ & $LQ\bar{D}$ & $\bar{U}\bar{D}\bar{D}$ & $QQQL$ & $\bar{U}\bar{U}\bar{D}\bar{E}$ & $QQQH_d$ & $Q\bar{U}\bar{E}H_d$ & $LH_uH_dH_u$ \\
    \hline\hline
    \endfirsthead
    
    \hline
    Case & $H_dH_u$ & $LH_u$ & $LL\bar{E}$ & $LQ\bar{D}$ & $\bar{U}\bar{D}\bar{D}$ & $QQQL$ & $\bar{U}\bar{U}\bar{D}\bar{E}$ & $QQQH_d$ & $Q\bar{U}\bar{E}H_d$ & $LH_uH_dH_u$ \\
    \hline\hline
    \endhead
    (1), (51), (58), (114) & \checkmark & \checkmark & \checkmark & \checkmark & \checkmark & \checkmark & \checkmark & \checkmark & \checkmark & \checkmark \\
    \hline
    (2), (3), (52), (53), (56), (57), (115), (116) & \checkmark &  &  &  & \checkmark &  &  & \checkmark &   &  \\
    \hline
    (4), (7), (8), (24), (27), (45), (48), (67), (70), (90), (93), (107), (110), (113) &  &  & \checkmark & \checkmark & \checkmark &  &  &  &  &  \checkmark \\
    \hline
    (5), (10), (46), (49), (68), (71), (109), (112) &  & \checkmark &  &  & \checkmark & \checkmark & \checkmark &  & \checkmark &  \\
    \hline
    (6), (111) &  &  &  &  & \checkmark &  & \checkmark &  & \checkmark &  \\
    \hline
    (9), (108) &  &  &  &  & \checkmark &  &  &  &   &  \\
    \hline
    (11), (21), (34), (37), (38), (41), (76), (79), (80), (83), (94), (104) &  &  & \checkmark & \checkmark &  & \checkmark & \checkmark & \checkmark &  & \checkmark \\
    \hline
    (12), (23), (35), (39), (78), (82), (96), (105) &  & \checkmark &  &  &  &  &  & \checkmark & \checkmark &  \\
    \hline
    (13), (106) &  &  &  &  &  &  & \checkmark & \checkmark &  & \checkmark \\
    \hline
    \rowcolor{blue!20}
    (14), (17), (28), (31), (84), (87), (100), (103) & \checkmark & \checkmark & \checkmark & \checkmark &  &  &  &  & \checkmark & \checkmark \\
    \hline
    \rowcolor{red!20}
    (15), (30), (32), (85), (89), (102) & \checkmark &  &  &  &  & \checkmark & \checkmark &  &  &  \\
    \hline
    (16), (101) & \checkmark &  &  &  &  &  & \checkmark &  &  &   \\
    \hline
    (18), (97) &  &  & \checkmark & \checkmark &  &  & \checkmark &  &  & \checkmark \\
    \hline
    (19), (98) &  & \checkmark &  &  &  &  &  &  & \checkmark &   \\
    \hline
    (20), (99) &  &  &  &  &  & \checkmark & \checkmark &  &  &  \checkmark \\
    \hline
    (22), (95) &  &  &  &  &  &  &  & \checkmark &  &  \\
    \hline
    (25), (92) &  &  &  &  & \checkmark & \checkmark & \checkmark &  & \checkmark &  \\
    \hline
    (26), (91) &  & \checkmark &  &  & \checkmark &  & \checkmark &  & \checkmark &  \\
    \hline
    \rowcolor{green!20}
    (29), (88) & \checkmark &  &  &  &  &  &  &  &  &  \\
    \hline
    (33), (86) & \checkmark &  &  & \checkmark &  &  &  &  & \checkmark &  \\
    \hline
    (36), (40), (77), (81) &  &  &  &  &  &  &  & \checkmark & \checkmark &  \\
    \hline
    (42), (73) &  &  & \checkmark & \checkmark &  &  &  &  & \checkmark & \checkmark \\
    \hline
    \rowcolor{red!20}
    (43), (74) &  &  &  &  &  & \checkmark & \checkmark &  &  &   \\
    \hline
    (44), (75) &  & \checkmark &  & \checkmark &  &  &  &  & \checkmark &  \\
    \hline
    (47), (50), (69), (72) &  &  &  & \checkmark & \checkmark &  &  &  &  & \checkmark \\
    \hline
    (54), (55) & \checkmark & \checkmark & \checkmark & \checkmark & \checkmark & \checkmark &  & \checkmark & \checkmark & \checkmark \\
    \hline
    (59), (66) & \checkmark & \checkmark & \checkmark & \checkmark &  & \checkmark & \checkmark & \checkmark & \checkmark & \checkmark \\
    \hline
    (60), (61), (64), (65) & \checkmark &  &  &  &  &  &  & \checkmark &   &  \\
    \hline
    (62), (63) & \checkmark & \checkmark & \checkmark & \checkmark &  & \checkmark &  & \checkmark & \checkmark &  \checkmark \\
    \hline
\end{longtable}
\end{tiny}
\begin{table}[H]
  \centering
  \caption{Allowed operators in D-terms represented by the symbol $\checkmark$ for $M=4$.}
  \label{tab:allowed-op-M=4D}
  \footnotesize
  \begin{tabular}{|p{4.0cm}  |c |c |c |c |}
    \hline
     &  \hspace{-.1cm} $\bar U \bar D^* \bar E$ \hspace{-.1cm} & \hspace{-.1cm}$H^*_u H_d \bar E$\hspace{-.1cm} &\hspace{-.1cm} $Q \bar U L^*$ \hspace{-.1cm} &\hspace{-.1cm} $QQ\bar D^*$ \hspace{-.1cm}  \\
    \hline
    (1), (12), (23), (35), (36), (39), (40), (51), (58), (62), (63), (77), (78), (81), (82), (96), (105), (114) & \checkmark & \checkmark & \checkmark & \checkmark \\ \hline
(2), (3), (11), (13), (21), (22), (34), (37), (38), (41), (52), (53), (56), (57), (60), (61), (64), (65), (76), (79), (80), (83), (94), (95), (104), (106), (115), (116) &  &  &  & \checkmark \\ \hline
(4), (7), (8), (9), \colorbox{red!20}{(15)}, (16), (18), (20), (24), (27), \colorbox{green!20}{(29)}, \colorbox{red!20}{(30)}, \colorbox{red!20}{(32)}, \colorbox{red!20}{(43)}, (45), (47), (48), (50), (67), (69), (70), (72), \colorbox{red!20}{(74)}, \colorbox{red!20}{(85)}, \colorbox{green!20}{(88)}, \colorbox{red!20}{(89)}, (90), (93), (97), (99), (101), \colorbox{red!20}{(102)}, (107), (108), (110), (113) &  &  &  &  \\ \hline
(5), (10), \colorbox{blue!20}{(14)}, \colorbox{blue!20}{(17}), (19), (26), \colorbox{blue!20}{(28)}, \colorbox{blue!20}{(31)}, (44), (46), (49), (68), (71), (75), \colorbox{blue!20}{(84)}, \colorbox{blue!20}{(87)}, (91), (98), \colorbox{blue!20}{(100)}, \colorbox{blue!20}{(103)}, (109), (112) & \checkmark & \checkmark & \checkmark &  \\ \hline
(6), (25), (92), (111) & \checkmark & \checkmark &  &  \\\hline
(33), (42), (73), (86) & \checkmark &  & \checkmark &  \\\hline
(54), (55), (59), (66) &  & \checkmark & \checkmark & \checkmark \\
    \hline
  \end{tabular}
\end{table}

Similarly, we can study the matter symmetries for $M=5$. Results are shown in Appendix~\ref{app}.

Judging from the patterns of allowed operators, we have various types of symmetries.
There exist the selection rules, which allow all of the operators shown in tables.
All symmetries allow the Weinberg operator $LH_uLH_u$.
Thus, we omitted the column for $LH_uLH_u$.
The reason why $LH_uLH_u$ is always allowed is as follows.
The product of the classes $[g^k][g^k]$ corresponding to $LL$ always includes $[g^0]$, and the product of the classes corresponding to $H_uH_u$ always includes $[g^0]$.
Therefore, the Weinberg operator $LH_uLH_u$ is always allowed.
In Ref.~\cite{Dreiner:2005rd}, several $\mathbb{Z}_N$ symmetries were studied.
Those symmetries except $R_2$, $B_3$, and $P_6$ forbid the Weinberg operators $LH_uLH_u$.
Our selection rules can not realize the same results as such symmetries forbidding $LH_uLH_u$, which are $R_3$, $L_3$, $R_3^2L_3$, $R_6$, $R^3_6L^2_6$, and $R_6L_6^2$ in the notation of Ref.~\cite{Dreiner:2005rd}.

On the other hand, the $\mu$-term is not always allowed.
Indeed, the $\mu$-term is forbidden when the classes of $H_u$ and $H_d$ are different from each other.
When the $\mu$-term is not allowed, we have to introduce the singlet chiral field $S$ with a proper class such that the three-point coupling $SH_uH_d$ is allowed as the next-to-minimal supersymmetric standard model.

Obviously, the selection rules for $M=4$ have more varieties.
There are the selection rules, which allow the same operators as 
$R_2$, $B_3$, and $P_6$.
These rows are colored in red, blue, and green.
The assignments (43) and (74) allow the same operators except $H_uH_d$ as $R_2$ as shown in Tables \ref{tab:allowed-op-M=4} and \ref{tab:allowed-op-M=4D}.
We also color the row in red. The assignments (2), (3), (52), (53), (56), (57), (115), and (116) almost correspond to the discrete $\mathbb{Z}_3$ symmetry associated with the lepton number ($L_3$ in Ref.~\cite{Ibanez:1991pr}), but they allow the Weinberg operators $LH_u LH_u$.

The strongest selection rules appear in the assignment to realize effectively $P_6$, i.e., the assignments (29) and (88) for $M=4$.
The next strongest rules appear in the assignments (16) and (101) for $M=4$, and they forbid all the dimension 4 operators and allow only the $\bar U \bar U \bar D \bar E$ as the dimension 5 operator in addition to $LH_uLH_u$.

If we assign the fields to $[g^0]$ and $[g^2]$ for $M=4$, the $\mathbb{Z}_2$ symmetries of $\mathbb{Z}_4$ remains. Such assignments are shown by the asterisk symbol $*$ in the last column of Table \ref{tab:assign-M=4}. 
In addition, the multiplication rules for $N=4$ includes the $\mathbb{Z}_2$ symmetry checked by the symbol $\checkmark$ in the last column of Table \ref{tab:assign-M=4}, which originates from the Ising fusion rule~(\ref{eq:Ising}). 
If we assign the fields to only two classes $[g^0]$ and $[g^1]$, or 
$[g^1]$ and $[g^2]$, their coupling selection rules are just the $\mathbb{Z}_2$ symmetry associated with $[g^1]\rightarrow -[g^1]$. 
For example, in the assignment (32), the Higgs pair $H_u$ and $H_d$ corresponds to $[g^0]$ and the other fields have $[g^1]$.
That is, the Higgs pair is $\mathbb{Z}_2$ even and the others are 
$\mathbb{Z}_2$ odd.
That is nothing but $R$-parity $R_2$.
In general, our selection rules are different from the selection rules due to only the $\mathbb{Z}_2$ symmetry when we assign the fields to three classes, $[g^0]$, $[g^1]$, and $[g^2]$.
However, our selection rule works effectively as the $\mathbb{Z}_2$ symmetry by a combination of the $SU(3) \times SU(2) \times U(1)_Y$ gauge symmetry even when we assign the fields to three classes, $[g^0]$, $[g^1]$, and $[g^2]$.
Table \ref{tab:R2-Z2} shows such examples, which behave effectively as the $\mathbb{Z}_2$ symmetry and allow the same operators as $R_2$.
In these assignments, we replace $[g^0]$ and $[g^2]$ by $\mathbb{Z}_2$ even and $[g^1]$ by $\mathbb{Z}_2$ odd.
Effectively, only these selection rules determine the allowed couplings.

\begin{table}[H]
    \centering
    \caption{$\mathbb{Z}_2$ charges which can lead the same allowed operators as $R_2$. We replace $[g^0]$ and $[g^2]$ by $\mathbb{Z}_2$ even and $[g^1]$ by $\mathbb{Z}_2$ odd.}
    \label{tab:R2-Z2}
    \begin{tabular}{|c|c|c|c|c|c|c|c|}
    \hline
      Case & $H_u$ & $H_d$ & $Q$ & $\bar{U}$ & $\bar{D}$ & $L$ & $\bar{E}$  \\ \hline
       (32), (43), (74), (85) & 0&0&1&1&1&1&1  \\ \hline
       (15), (102) &1&1&0&1&1&0&1 \\ \hline
        \end{tabular}
\end{table}

Similarly, a combination of only the $\mathbb{Z}_2$ symmetry of our selection rules and the $SU(3) \times SU(2) \times U(1)_Y$ gauge symmetry can lead to the same allowed operators as $B_3$.
Table \ref{tab:B3-Z2} shows such assignments.
In addition to these assignments in Tables~\ref{tab:R2-Z2} and \ref{tab:B3-Z2}, there are other assignments, whose selection rules behave effectively as the $\mathbb{Z}_2$ symmetry.
These assignments are shown by the check symbol $\checkmark$ in the last column of Table \ref{tab:assign-M=4}.

\begin{table}[H]
    \centering
    \caption{$\mathbb{Z}_2$ charges which can lead the same allowed operators as $B_3$. We replace $[g^0]$ and $[g^2]$ by $\mathbb{Z}_2$ even and $[g^1]$ by $\mathbb{Z}_2$ odd.}
    \label{tab:B3-Z2}
    \begin{tabular}{|c|c|c|c|c|c|c|c|}
    \hline
      Case & $H_u$ & $H_d$ & $Q$ & $\bar{U}$ & $\bar{D}$ & $L$ & $\bar{E}$  \\ \hline
       (31), (84) & 0&0&1&1&1&0&0  \\ \hline
       (14), (17), (100), (103)  &1&1&0&1&1&1&0 \\ \hline
        \end{tabular}
\end{table}

It is also interesting whether the assignments are consistent with 
$SU(5)$, $SU(4) \times SU(2) \times SU(2)$, and $SO(10)$ grand unified theories.
The columns written as SU5, PS SO10 in Tables show the consistency with 
$SU(5)$, $SU(4) \times SU(2) \times SU(2)$, and $SO(10)$ grand unified theories.

\section{Conclusions}
\label{sec:con}

We clarified the phenomenological role of non-invertible selection rules in the supersymmetric standard model, where the matter fields are assigned to the elements of fusion algebras in a family-independent way. In particular, we focused on the non-invertible selection rules originating from $\mathbb{Z}_2$ gauging of $\mathbb{Z}_M$ symmetries such as the Fibonacci fusion rule for $M = 3$ and Ising fusion rule $M = 4$.

So far, matter $\mathbb{Z}_N$ symmetries have been classified in the context of the supersymmetric standard model, leading to $R$-parity, baryon triality, proton hexality, and so on. By contrast, our finding results are summarized as follows:
\begin{itemize}
    \item The Weinberg operator $LH_u LH_u$ is always allowed due to the existence of the identity as a multiplication of the same class $[g^k][g^k]$, where $L$ and/or $H_u$ are assigned to $[g^k]$. This result is different from the conventional $\mathbb{Z}_N$ symmetries in the supersymmetric standard model~\cite{Dreiner:2005rd}. 

    \item In the case of the Ising fusion rule realized for $M=4$, one can obtain $R$-parity, baryon triality, and proton hexality in the effective action. In particular, the $R$-parity arises as a consequence of a remnant $\mathbb{Z}_2$ symmetry of the Ising fusion algebra which holds at all-loop order. Together with the selection rule of the Standard Model gauge symmetry, the baryon triality, and proton hexality can be realized, although we have not explicitly imposed $\mathbb{Z}_3$ and $\mathbb{Z}_6$ symmetries. Such a $\mathbb{Z}_2$ symmetry protects the stability of lightest supersymmetric particle, i.e., a candidate of dark matter.

    \item The non-invertible selection rules lead to the different results from the conventional $\mathbb{Z}_N$ symmetries as shown in Tables~\ref{tab:allowed-op-M=4}, but some of them can be uplifted to $SU(4) \times SU(2) \times SU(2)$, $SU(5)$, and $SO(10)$ grand unified theories. It would be interesting to construct explicit grand unified theories with the non-invertible selection rules.

\end{itemize}

We have studied family-independent matter symmetries. 
On the other hand, in Refs.~\cite{Kobayashi:2024cvp,Kobayashi:2025znw,Kobayashi:2025ldi}, 
it were shown that 
interesting mass textures can be derived by assigning family-dependent classes in $\mathbb{Z}_2$ gauging of $\mathbb{Z}_M$ symmetries. 
It is important to combine these family-independent and dependent classes. The simplest combination would be a kind of their direct product. Also we can discuss non-trivial combinations. 
We would study such topics in the supersymmetric standard model and grand unified theories in the future.

We have derived various selection rules. Allowed operators depend on these selection rules. 
Different selection rules may lead to different phenomenology and cosmology. These operators have experimental constraints, e.g. by the proton decay. (See e.g. Ref.~\cite{Barbier:2004ez}.) The supersymmetric standard model with only Yukawa terms have several flat directions, which are interesting from particle phenomenology and cosmology. (See e.g. Ref.~\cite{Enqvist:2003gh}.) Additional operators would lift up such flat directions. Furthermore, soft supersymmetry breaking terms corresponding to allowed operators may change the vacuum structure such as color and/or charge breaking and unbounded-from-below directions.  It would be important to study details of phenomenological and cosmological aspects of our selection rules. We leave it as our future work.

\acknowledgments

This work was supported by JSPS KAKENHI Grant Numbers JP23K03375 (T.K.) and JP25H01539 (H.O.).

\appendix

\section{$M=5$}
\label{app}

Similarly, we can study the coupling selection rules due to $\mathbb{Z}_2$ gauging of $\mathbb{Z}_5$ symmetry.
There are three classes, $[g^0]$, $[g^1]$, and $[g^2]$.
Table \ref{tab:assign-M=5} shows all the possible assignments of classes, which allow Yukawa couplings (\ref{eq:YUkawa}).
Tables \ref{tab:allowed-op-M=5} and \ref{tab:allowed-op-M=5D} show allowed operators by each assignment.

\begin{longtable}{|l |c |c |c |c |c |c |c |c |c |c |}
    \caption{Possible assignments to allow Yukawa couplings for $M=5$, where the symbol $\checkmark$ denotes the consistent assignments of matter fields under $SU(5)$, $SU(4) \times SU(2) \times SU(2)$, and $SO(10)$ grand unified theories, represented by SU5, PS, and SO10 respectively.}
    \label{tab:assign-M=5} \\
    \hline
    Case & $Q$ & $H_u$ & $H_d$ & $\bar{U}$ & $\bar{D}$ & $\bar{E}$ & $L$ & SU5 & PS & SO10 \\
    \hline
    \endfirsthead
    \hline
    Case & $Q$ & $H_u$ & $H_d$ & $\bar{U}$ & $\bar{D}$ & $\bar{E}$ & $L$ & SU5 & PS & SO10 \\
    \hline
    \endhead
    ${M=5}$ (1) & 0 & 0 & 0 & 0 & 0 & 0 & 0 & \checkmark & \checkmark & \checkmark \\
    \hline
    ${M=5}$ (2) & 0 & 0 & 0 & 0 & 0 & 1 & 1 &  &  &  \\
    \hline
    ${M=5}$ (3) & 0 & 0 & 0 & 0 & 0 & 2 & 2 &  &  &  \\
    \hline
    ${M=5}$ (4) & 0 & 0 & 1 & 0 & 1 & 0 & 1 & \checkmark &  &  \\
    \hline
    ${M=5}$ (5) & 0 & 0 & 1 & 0 & 1 & 1 & 0 &  &  &  \\
    \hline
    ${M=5}$ (6) & 0 & 0 & 1 & 0 & 1 & 1 & 2 &  &  &  \\
    \hline
    ${M=5}$ (7) & 0 & 0 & 1 & 0 & 1 & 2 & 1 &  &  &  \\
    \hline
    ${M=5}$ (8) & 0 & 0 & 1 & 0 & 1 & 2 & 2 &  &  &  \\
    \hline
    ${M=5}$ (9) & 0 & 0 & 2 & 0 & 2 & 0 & 2 & \checkmark &  &  \\
    \hline
    ${M=5}$ (10) & 0 & 0 & 2 & 0 & 2 & 1 & 1 &  &  &  \\
    \hline
    ${M=5}$ (11) & 0 & 0 & 2 & 0 & 2 & 1 & 2 &  &  &  \\
    \hline
    ${M=5}$ (12) & 0 & 0 & 2 & 0 & 2 & 2 & 0 &  &  &  \\
    \hline
    ${M=5}$ (13) & 0 & 0 & 2 & 0 & 2 & 2 & 1 &  &  &  \\
    \hline
    ${M=5}$ (14) & 0 & 1 & 0 & 1 & 0 & 0 & 0 &  &  &  \\
    \hline
    ${M=5}$ (15) & 0 & 1 & 0 & 1 & 0 & 1 & 1 &  &  &  \\
    \hline
    ${M=5}$ (16) & 0 & 1 & 0 & 1 & 0 & 2 & 2 &  &  &  \\
    \hline
    \rowcolor{blue!20}${M=5}$ (17) & 0 & 1 & 1 & 1 & 1 & 0 & 1 &  &  &  \\
    \hline
    \rowcolor{red!20}${M=5}$ (18) & 0 & 1 & 1 & 1 & 1 & 1 & 0 &  & \checkmark &  \\
    \hline
    ${M=5}$ (19) & 0 & 1 & 1 & 1 & 1 & 1 & 2 &  &  &  \\
    \hline
    ${M=5}$ (20) & 0 & 1 & 1 & 1 & 1 & 2 & 1 &  &  &  \\
    \hline
    ${M=5}$ (21) & 0 & 1 & 1 & 1 & 1 & 2 & 2 &  &  &  \\
    \hline
    ${M=5}$ (22) & 0 & 1 & 2 & 1 & 2 & 0 & 2 &  &  &  \\
    \hline
    ${M=5}$ (23) & 0 & 1 & 2 & 1 & 2 & 1 & 1 &  &  &  \\
    \hline
    ${M=5}$ (24) & 0 & 1 & 2 & 1 & 2 & 1 & 2 &  &  &  \\
    \hline
    ${M=5}$ (25) & 0 & 1 & 2 & 1 & 2 & 2 & 0 &  &  &  \\
    \hline
    ${M=5}$ (26) & 0 & 1 & 2 & 1 & 2 & 2 & 1 &  &  &  \\
    \hline
    ${M=5}$ (27) & 0 & 2 & 0 & 2 & 0 & 0 & 0 &  &  &  \\
    \hline
    ${M=5}$ (28) & 0 & 2 & 0 & 2 & 0 & 1 & 1 &  &  &  \\
    \hline
    ${M=5}$ (29) & 0 & 2 & 0 & 2 & 0 & 2 & 2 &  &  &  \\
    \hline
    ${M=5}$ (30) & 0 & 2 & 1 & 2 & 1 & 0 & 1 &  &  &  \\
    \hline
    ${M=5}$ (31) & 0 & 2 & 1 & 2 & 1 & 1 & 0 &  &  &  \\
    \hline
    ${M=5}$ (32) & 0 & 2 & 1 & 2 & 1 & 1 & 2 &  &  &  \\
    \hline
    ${M=5}$ (33) & 0 & 2 & 1 & 2 & 1 & 2 & 1 &  &  &  \\
    \hline
    ${M=5}$ (34) & 0 & 2 & 1 & 2 & 1 & 2 & 2 &  &  &  \\
    \hline
    \rowcolor{blue!20}${M=5}$ (35) & 0 & 2 & 2 & 2 & 2 & 0 & 2 &  &  &  \\
    \hline
    ${M=5}$ (36) & 0 & 2 & 2 & 2 & 2 & 1 & 1 &  &  &  \\
    \hline
    ${M=5}$ (37) & 0 & 2 & 2 & 2 & 2 & 1 & 2 &  &  &  \\
    \hline
    \rowcolor{red!20}${M=5}$ (38) & 0 & 2 & 2 & 2 & 2 & 2 & 0 &  & \checkmark &  \\
    \hline
    ${M=5}$ (39) & 0 & 2 & 2 & 2 & 2 & 2 & 1 &  &  &  \\
    \hline
    \rowcolor{blue!20}${M=5}$ (40) & 1 & 0 & 0 & 1 & 1 & 0 & 0 &  &  &  \\
    \hline
    \rowcolor{red!20}${M=5}$ (41) & 1 & 0 & 0 & 1 & 1 & 1 & 1 & \checkmark & \checkmark & \checkmark \\
    \hline
    ${M=5}$ (42) & 1 & 0 & 0 & 1 & 1 & 2 & 2 &  &  &  \\
    \hline
    ${M=5}$ (43) & 1 & 0 & 1 & 1 & 0 & 0 & 1 &  &  &  \\
    \hline
    ${M=5}$ (44) & 1 & 0 & 1 & 1 & 0 & 1 & 0 & \checkmark &  &  \\
    \hline
    ${M=5}$ (45) & 1 & 0 & 1 & 1 & 0 & 1 & 2 &  &  &  \\
    \hline
    ${M=5}$ (46) & 1 & 0 & 1 & 1 & 0 & 2 & 1 &  &  &  \\
    \hline
    ${M=5}$ (47) & 1 & 0 & 1 & 1 & 0 & 2 & 2 &  &  &  \\
    \hline
    ${M=5}$ (48) & 1 & 0 & 1 & 1 & 2 & 0 & 1 &  &  &  \\
    \hline
    ${M=5}$ (49) & 1 & 0 & 1 & 1 & 2 & 1 & 0 &  &  &  \\
    \hline
    ${M=5}$ (50) & 1 & 0 & 1 & 1 & 2 & 1 & 2 & \checkmark &  &  \\
    \hline
    ${M=5}$ (51) & 1 & 0 & 1 & 1 & 2 & 2 & 1 &  &  &  \\
    \hline
    ${M=5}$ (52) & 1 & 0 & 1 & 1 & 2 & 2 & 2 &  &  &  \\
    \hline
    ${M=5}$ (53) & 1 & 0 & 2 & 1 & 1 & 0 & 2 &  &  &  \\
    \hline
    ${M=5}$ (54) & 1 & 0 & 2 & 1 & 1 & 1 & 1 & \checkmark & \checkmark &  \\
    \hline
    ${M=5}$ (55) & 1 & 0 & 2 & 1 & 1 & 1 & 2 &  &  &  \\
    \hline
    ${M=5}$ (56) & 1 & 0 & 2 & 1 & 1 & 2 & 0 &  &  &  \\
    \hline
    ${M=5}$ (57) & 1 & 0 & 2 & 1 & 1 & 2 & 1 &  &  &  \\
    \hline
    ${M=5}$ (58) & 1 & 0 & 2 & 1 & 2 & 0 & 2 &  &  &  \\
    \hline
    ${M=5}$ (59) & 1 & 0 & 2 & 1 & 2 & 1 & 1 &  &  &  \\
    \hline
    ${M=5}$ (60) & 1 & 0 & 2 & 1 & 2 & 1 & 2 & \checkmark &  &  \\
    \hline
    ${M=5}$ (61) & 1 & 0 & 2 & 1 & 2 & 2 & 0 &  &  &  \\
    \hline
    ${M=5}$ (62) & 1 & 0 & 2 & 1 & 2 & 2 & 1 &  &  &  \\
    \hline
    ${M=5}$ (63) & 1 & 1 & 0 & 0 & 1 & 0 & 0 &  &  &  \\
    \hline
    ${M=5}$ (64) & 1 & 1 & 0 & 0 & 1 & 1 & 1 &  &  &  \\
    \hline
    ${M=5}$ (65) & 1 & 1 & 0 & 0 & 1 & 2 & 2 &  &  &  \\
    \hline
    ${M=5}$ (66) & 1 & 1 & 0 & 2 & 1 & 0 & 0 &  &  &  \\
    \hline
    ${M=5}$ (67) & 1 & 1 & 0 & 2 & 1 & 1 & 1 &  &  &  \\
    \hline
    ${M=5}$ (68) & 1 & 1 & 0 & 2 & 1 & 2 & 2 &  &  &  \\
    \hline
    ${M=5}$ (69) & 1 & 1 & 1 & 0 & 0 & 0 & 1 &  & \checkmark &  \\
    \hline
    ${M=5}$ (70) & 1 & 1 & 1 & 0 & 0 & 1 & 0 &  &  &  \\
    \hline
    ${M=5}$ (71) & 1 & 1 & 1 & 0 & 0 & 1 & 2 &  &  &  \\
    \hline
    ${M=5}$ (72) & 1 & 1 & 1 & 0 & 0 & 2 & 1 &  &  &  \\
    \hline
    ${M=5}$ (73) & 1 & 1 & 1 & 0 & 0 & 2 & 2 &  &  &  \\
    \hline
    ${M=5}$ (74) & 1 & 1 & 1 & 0 & 2 & 0 & 1 &  &  &  \\
    \hline
    ${M=5}$ (75) & 1 & 1 & 1 & 0 & 2 & 1 & 0 &  &  &  \\
    \hline
    ${M=5}$ (76) & 1 & 1 & 1 & 0 & 2 & 1 & 2 &  &  &  \\
    \hline
    ${M=5}$ (77) & 1 & 1 & 1 & 0 & 2 & 2 & 1 &  &  &  \\
    \hline
    ${M=5}$ (78) & 1 & 1 & 1 & 0 & 2 & 2 & 2 &  &  &  \\
    \hline
    ${M=5}$ (79) & 1 & 1 & 1 & 2 & 0 & 0 & 1 &  &  &  \\
    \hline
    ${M=5}$ (80) & 1 & 1 & 1 & 2 & 0 & 1 & 0 &  &  &  \\
    \hline
    ${M=5}$ (81) & 1 & 1 & 1 & 2 & 0 & 1 & 2 &  &  &  \\
    \hline
    ${M=5}$ (82) & 1 & 1 & 1 & 2 & 0 & 2 & 1 &  &  &  \\
    \hline
    ${M=5}$ (83) & 1 & 1 & 1 & 2 & 0 & 2 & 2 &  &  &  \\
    \hline
    ${M=5}$ (84) & 1 & 1 & 1 & 2 & 2 & 0 & 1 &  &  &  \\
    \hline
    ${M=5}$ (85) & 1 & 1 & 1 & 2 & 2 & 1 & 0 &  &  &  \\
    \hline
    ${M=5}$ (86) & 1 & 1 & 1 & 2 & 2 & 1 & 2 &  &  &  \\
    \hline
    ${M=5}$ (87) & 1 & 1 & 1 & 2 & 2 & 2 & 1 &  & \checkmark &  \\
    \hline
    ${M=5}$ (88) & 1 & 1 & 1 & 2 & 2 & 2 & 2 &  &  &  \\
    \hline
    ${M=5}$ (89) & 1 & 1 & 2 & 0 & 1 & 0 & 2 &  &  &  \\
    \hline
    ${M=5}$ (90) & 1 & 1 & 2 & 0 & 1 & 1 & 1 &  &  &  \\
    \hline
    ${M=5}$ (91) & 1 & 1 & 2 & 0 & 1 & 1 & 2 &  &  &  \\
    \hline
    ${M=5}$ (92) & 1 & 1 & 2 & 0 & 1 & 2 & 0 &  &  &  \\
    \hline
    ${M=5}$ (93) & 1 & 1 & 2 & 0 & 1 & 2 & 1 &  &  &  \\
    \hline
    ${M=5}$ (94) & 1 & 1 & 2 & 0 & 2 & 0 & 2 &  &  &  \\
    \hline
    ${M=5}$ (95) & 1 & 1 & 2 & 0 & 2 & 1 & 1 &  &  &  \\
    \hline
    ${M=5}$ (96) & 1 & 1 & 2 & 0 & 2 & 1 & 2 &  &  &  \\
    \hline
    ${M=5}$ (97) & 1 & 1 & 2 & 0 & 2 & 2 & 0 &  &  &  \\
    \hline
    ${M=5}$ (98) & 1 & 1 & 2 & 0 & 2 & 2 & 1 &  &  &  \\
    \hline
    ${M=5}$ (99) & 1 & 1 & 2 & 2 & 1 & 0 & 2 &  &  &  \\
    \hline
    ${M=5}$ (100) & 1 & 1 & 2 & 2 & 1 & 1 & 1 &  &  &  \\
    \hline
    ${M=5}$ (101) & 1 & 1 & 2 & 2 & 1 & 1 & 2 &  &  &  \\
    \hline
    ${M=5}$ (102) & 1 & 1 & 2 & 2 & 1 & 2 & 0 &  &  &  \\
    \hline
    ${M=5}$ (103) & 1 & 1 & 2 & 2 & 1 & 2 & 1 &  &  &  \\
    \hline
    ${M=5}$ (104) & 1 & 1 & 2 & 2 & 2 & 0 & 2 &  &  &  \\
    \hline
    ${M=5}$ (105) & 1 & 1 & 2 & 2 & 2 & 1 & 1 &  &  &  \\
    \hline
    ${M=5}$ (106) & 1 & 1 & 2 & 2 & 2 & 1 & 2 &  &  &  \\
    \hline
    ${M=5}$ (107) & 1 & 1 & 2 & 2 & 2 & 2 & 0 &  &  &  \\
    \hline
    ${M=5}$ (108) & 1 & 1 & 2 & 2 & 2 & 2 & 1 &  & \checkmark &  \\
    \hline
    ${M=5}$ (109) & 1 & 2 & 0 & 1 & 1 & 0 & 0 &  &  &  \\
    \hline
    ${M=5}$ (110) & 1 & 2 & 0 & 1 & 1 & 1 & 1 & \checkmark & \checkmark &  \\
    \hline
    ${M=5}$ (111) & 1 & 2 & 0 & 1 & 1 & 2 & 2 &  &  &  \\
    \hline
    ${M=5}$ (112) & 1 & 2 & 0 & 2 & 1 & 0 & 0 &  &  &  \\
    \hline
    ${M=5}$ (113) & 1 & 2 & 0 & 2 & 1 & 1 & 1 &  &  &  \\
    \hline
    ${M=5}$ (114) & 1 & 2 & 0 & 2 & 1 & 2 & 2 &  &  &  \\
    \hline
    ${M=5}$ (115) & 1 & 2 & 1 & 1 & 0 & 0 & 1 &  &  &  \\
    \hline
    ${M=5}$ (116) & 1 & 2 & 1 & 1 & 0 & 1 & 0 & \checkmark &  &  \\
    \hline
    ${M=5}$ (117) & 1 & 2 & 1 & 1 & 0 & 1 & 2 &  &  &  \\
    \hline
    ${M=5}$ (118) & 1 & 2 & 1 & 1 & 0 & 2 & 1 &  &  &  \\
    \hline
    ${M=5}$ (119) & 1 & 2 & 1 & 1 & 0 & 2 & 2 &  &  &  \\
    \hline
    ${M=5}$ (120) & 1 & 2 & 1 & 1 & 2 & 0 & 1 &  &  &  \\
    \hline
    ${M=5}$ (121) & 1 & 2 & 1 & 1 & 2 & 1 & 0 &  &  &  \\
    \hline
    ${M=5}$ (122) & 1 & 2 & 1 & 1 & 2 & 1 & 2 & \checkmark &  &  \\
    \hline
    ${M=5}$ (123) & 1 & 2 & 1 & 1 & 2 & 2 & 1 &  &  &  \\
    \hline
    ${M=5}$ (124) & 1 & 2 & 1 & 1 & 2 & 2 & 2 &  &  &  \\
    \hline
    ${M=5}$ (125) & 1 & 2 & 1 & 2 & 0 & 0 & 1 &  &  &  \\
    \hline
    ${M=5}$ (126) & 1 & 2 & 1 & 2 & 0 & 1 & 0 &  &  &  \\
    \hline
    ${M=5}$ (127) & 1 & 2 & 1 & 2 & 0 & 1 & 2 &  &  &  \\
    \hline
    ${M=5}$ (128) & 1 & 2 & 1 & 2 & 0 & 2 & 1 &  &  &  \\
    \hline
    ${M=5}$ (129) & 1 & 2 & 1 & 2 & 0 & 2 & 2 &  &  &  \\
    \hline
    ${M=5}$ (130) & 1 & 2 & 1 & 2 & 2 & 0 & 1 &  &  &  \\
    \hline
    ${M=5}$ (131) & 1 & 2 & 1 & 2 & 2 & 1 & 0 &  &  &  \\
    \hline
    ${M=5}$ (132) & 1 & 2 & 1 & 2 & 2 & 1 & 2 &  &  &  \\
    \hline
    ${M=5}$ (133) & 1 & 2 & 1 & 2 & 2 & 2 & 1 &  & \checkmark &  \\
    \hline
    ${M=5}$ (134) & 1 & 2 & 1 & 2 & 2 & 2 & 2 &  &  &  \\
    \hline
    ${M=5}$ (135) & 1 & 2 & 2 & 1 & 1 & 0 & 2 &  &  &  \\
    \hline
    ${M=5}$ (136) & 1 & 2 & 2 & 1 & 1 & 1 & 1 & \checkmark & \checkmark & \checkmark \\
    \hline
    ${M=5}$ (137) & 1 & 2 & 2 & 1 & 1 & 1 & 2 &  &  &  \\
    \hline
    ${M=5}$ (138) & 1 & 2 & 2 & 1 & 1 & 2 & 0 &  &  &  \\
    \hline
    ${M=5}$ (139) & 1 & 2 & 2 & 1 & 1 & 2 & 1 &  &  &  \\
    \hline
    ${M=5}$ (140) & 1 & 2 & 2 & 1 & 2 & 0 & 2 &  &  &  \\
    \hline
    ${M=5}$ (141) & 1 & 2 & 2 & 1 & 2 & 1 & 1 &  &  &  \\
    \hline
    ${M=5}$ (142) & 1 & 2 & 2 & 1 & 2 & 1 & 2 & \checkmark &  &  \\
    \hline
    ${M=5}$ (143) & 1 & 2 & 2 & 1 & 2 & 2 & 0 &  &  &  \\
    \hline
    ${M=5}$ (144) & 1 & 2 & 2 & 1 & 2 & 2 & 1 &  &  &  \\
    \hline
    ${M=5}$ (145) & 1 & 2 & 2 & 2 & 1 & 0 & 2 &  &  &  \\
    \hline
    ${M=5}$ (146) & 1 & 2 & 2 & 2 & 1 & 1 & 1 &  &  &  \\
    \hline
    ${M=5}$ (147) & 1 & 2 & 2 & 2 & 1 & 1 & 2 &  &  &  \\
    \hline
    ${M=5}$ (148) & 1 & 2 & 2 & 2 & 1 & 2 & 0 &  &  &  \\
    \hline
    ${M=5}$ (149) & 1 & 2 & 2 & 2 & 1 & 2 & 1 &  &  &  \\
    \hline
    ${M=5}$ (150) & 1 & 2 & 2 & 2 & 2 & 0 & 2 &  &  &  \\
    \hline
    ${M=5}$ (151) & 1 & 2 & 2 & 2 & 2 & 1 & 1 &  &  &  \\
    \hline
    ${M=5}$ (152) & 1 & 2 & 2 & 2 & 2 & 1 & 2 &  &  &  \\
    \hline
    ${M=5}$ (153) & 1 & 2 & 2 & 2 & 2 & 2 & 0 &  &  &  \\
    \hline
    ${M=5}$ (154) & 1 & 2 & 2 & 2 & 2 & 2 & 1 &  & \checkmark &  \\
    \hline
    \rowcolor{blue!20}${M=5}$ (155) & 2 & 0 & 0 & 2 & 2 & 0 & 0 &  &  &  \\
    \hline
    ${M=5}$ (156) & 2 & 0 & 0 & 2 & 2 & 1 & 1 &  &  &  \\
    \hline
    \rowcolor{red!20}${M=5}$ (157) & 2 & 0 & 0 & 2 & 2 & 2 & 2 & \checkmark & \checkmark & \checkmark \\
    \hline
    ${M=5}$ (158) & 2 & 0 & 1 & 2 & 1 & 0 & 1 &  &  &  \\
    \hline
    ${M=5}$ (159) & 2 & 0 & 1 & 2 & 1 & 1 & 0 &  &  &  \\
    \hline
    ${M=5}$ (160) & 2 & 0 & 1 & 2 & 1 & 1 & 2 &  &  &  \\
    \hline
    ${M=5}$ (161) & 2 & 0 & 1 & 2 & 1 & 2 & 1 & \checkmark &  &  \\
    \hline
    ${M=5}$ (162) & 2 & 0 & 1 & 2 & 1 & 2 & 2 &  &  &  \\
    \hline
    ${M=5}$ (163) & 2 & 0 & 1 & 2 & 2 & 0 & 1 &  &  &  \\
    \hline
    ${M=5}$ (164) & 2 & 0 & 1 & 2 & 2 & 1 & 0 &  &  &  \\
    \hline
    ${M=5}$ (165) & 2 & 0 & 1 & 2 & 2 & 1 & 2 &  &  &  \\
    \hline
    ${M=5}$ (166) & 2 & 0 & 1 & 2 & 2 & 2 & 1 &  &  &  \\
    \hline
    ${M=5}$ (167) & 2 & 0 & 1 & 2 & 2 & 2 & 2 & \checkmark & \checkmark &  \\
    \hline
    ${M=5}$ (168) & 2 & 0 & 2 & 2 & 0 & 0 & 2 &  &  &  \\
    \hline
    ${M=5}$ (169) & 2 & 0 & 2 & 2 & 0 & 1 & 1 &  &  &  \\
    \hline
    ${M=5}$ (170) & 2 & 0 & 2 & 2 & 0 & 1 & 2 &  &  &  \\
    \hline
    ${M=5}$ (171) & 2 & 0 & 2 & 2 & 0 & 2 & 0 & \checkmark &  &  \\
    \hline
    ${M=5}$ (172) & 2 & 0 & 2 & 2 & 0 & 2 & 1 &  &  &  \\
    \hline
    ${M=5}$ (173) & 2 & 0 & 2 & 2 & 1 & 0 & 2 &  &  &  \\
    \hline
    ${M=5}$ (174) & 2 & 0 & 2 & 2 & 1 & 1 & 1 &  &  &  \\
    \hline
    ${M=5}$ (175) & 2 & 0 & 2 & 2 & 1 & 1 & 2 &  &  &  \\
    \hline
    ${M=5}$ (176) & 2 & 0 & 2 & 2 & 1 & 2 & 0 &  &  &  \\
    \hline
    ${M=5}$ (177) & 2 & 0 & 2 & 2 & 1 & 2 & 1 & \checkmark &  &  \\
    \hline
    ${M=5}$ (178) & 2 & 1 & 0 & 1 & 2 & 0 & 0 &  &  &  \\
    \hline
    ${M=5}$ (179) & 2 & 1 & 0 & 1 & 2 & 1 & 1 &  &  &  \\
    \hline
    ${M=5}$ (180) & 2 & 1 & 0 & 1 & 2 & 2 & 2 &  &  &  \\
    \hline
    ${M=5}$ (181) & 2 & 1 & 0 & 2 & 2 & 0 & 0 &  &  &  \\
    \hline
    ${M=5}$ (182) & 2 & 1 & 0 & 2 & 2 & 1 & 1 &  &  &  \\
    \hline
    ${M=5}$ (183) & 2 & 1 & 0 & 2 & 2 & 2 & 2 & \checkmark & \checkmark &  \\
    \hline
    ${M=5}$ (184) & 2 & 1 & 1 & 1 & 1 & 0 & 1 &  &  &  \\
    \hline
    ${M=5}$ (185) & 2 & 1 & 1 & 1 & 1 & 1 & 0 &  &  &  \\
    \hline
    ${M=5}$ (186) & 2 & 1 & 1 & 1 & 1 & 1 & 2 &  & \checkmark &  \\
    \hline
    ${M=5}$ (187) & 2 & 1 & 1 & 1 & 1 & 2 & 1 &  &  &  \\
    \hline
    ${M=5}$ (188) & 2 & 1 & 1 & 1 & 1 & 2 & 2 &  &  &  \\
    \hline
    ${M=5}$ (189) & 2 & 1 & 1 & 1 & 2 & 0 & 1 &  &  &  \\
    \hline
    ${M=5}$ (190) & 2 & 1 & 1 & 1 & 2 & 1 & 0 &  &  &  \\
    \hline
    ${M=5}$ (191) & 2 & 1 & 1 & 1 & 2 & 1 & 2 &  &  &  \\
    \hline
    ${M=5}$ (192) & 2 & 1 & 1 & 1 & 2 & 2 & 1 &  &  &  \\
    \hline
    ${M=5}$ (193) & 2 & 1 & 1 & 1 & 2 & 2 & 2 &  &  &  \\
    \hline
    ${M=5}$ (194) & 2 & 1 & 1 & 2 & 1 & 0 & 1 &  &  &  \\
    \hline
    ${M=5}$ (195) & 2 & 1 & 1 & 2 & 1 & 1 & 0 &  &  &  \\
    \hline
    ${M=5}$ (196) & 2 & 1 & 1 & 2 & 1 & 1 & 2 &  &  &  \\
    \hline
    ${M=5}$ (197) & 2 & 1 & 1 & 2 & 1 & 2 & 1 & \checkmark &  &  \\
    \hline
    ${M=5}$ (198) & 2 & 1 & 1 & 2 & 1 & 2 & 2 &  &  &  \\
    \hline
    ${M=5}$ (199) & 2 & 1 & 1 & 2 & 2 & 0 & 1 &  &  &  \\
    \hline
    ${M=5}$ (200) & 2 & 1 & 1 & 2 & 2 & 1 & 0 &  &  &  \\
    \hline
    ${M=5}$ (201) & 2 & 1 & 1 & 2 & 2 & 1 & 2 &  &  &  \\
    \hline
    ${M=5}$ (202) & 2 & 1 & 1 & 2 & 2 & 2 & 1 &  &  &  \\
    \hline
    ${M=5}$ (203) & 2 & 1 & 1 & 2 & 2 & 2 & 2 & \checkmark & \checkmark & \checkmark \\
    \hline
    ${M=5}$ (204) & 2 & 1 & 2 & 1 & 0 & 0 & 2 &  &  &  \\
    \hline
    ${M=5}$ (205) & 2 & 1 & 2 & 1 & 0 & 1 & 1 &  &  &  \\
    \hline
    ${M=5}$ (206) & 2 & 1 & 2 & 1 & 0 & 1 & 2 &  &  &  \\
    \hline
    ${M=5}$ (207) & 2 & 1 & 2 & 1 & 0 & 2 & 0 &  &  &  \\
    \hline
    ${M=5}$ (208) & 2 & 1 & 2 & 1 & 0 & 2 & 1 &  &  &  \\
    \hline
    ${M=5}$ (209) & 2 & 1 & 2 & 1 & 1 & 0 & 2 &  &  &  \\
    \hline
    ${M=5}$ (210) & 2 & 1 & 2 & 1 & 1 & 1 & 1 &  &  &  \\
    \hline
    ${M=5}$ (211) & 2 & 1 & 2 & 1 & 1 & 1 & 2 &  & \checkmark &  \\
    \hline
    ${M=5}$ (212) & 2 & 1 & 2 & 1 & 1 & 2 & 0 &  &  &  \\
    \hline
    ${M=5}$ (213) & 2 & 1 & 2 & 1 & 1 & 2 & 1 &  &  &  \\
    \hline
    ${M=5}$ (214) & 2 & 1 & 2 & 2 & 0 & 0 & 2 &  &  &  \\
    \hline
    ${M=5}$ (215) & 2 & 1 & 2 & 2 & 0 & 1 & 1 &  &  &  \\
    \hline
    ${M=5}$ (216) & 2 & 1 & 2 & 2 & 0 & 1 & 2 &  &  &  \\
    \hline
    ${M=5}$ (217) & 2 & 1 & 2 & 2 & 0 & 2 & 0 & \checkmark &  &  \\
    \hline
    ${M=5}$ (218) & 2 & 1 & 2 & 2 & 0 & 2 & 1 &  &  &  \\
    \hline
    ${M=5}$ (219) & 2 & 1 & 2 & 2 & 1 & 0 & 2 &  &  &  \\
    \hline
    ${M=5}$ (220) & 2 & 1 & 2 & 2 & 1 & 1 & 1 &  &  &  \\
    \hline
    ${M=5}$ (221) & 2 & 1 & 2 & 2 & 1 & 1 & 2 &  &  &  \\
    \hline
    ${M=5}$ (222) & 2 & 1 & 2 & 2 & 1 & 2 & 0 &  &  &  \\
    \hline
    ${M=5}$ (223) & 2 & 1 & 2 & 2 & 1 & 2 & 1 & \checkmark &  &  \\
    \hline
    ${M=5}$ (224) & 2 & 2 & 0 & 0 & 2 & 0 & 0 &  &  &  \\
    \hline
    ${M=5}$ (225) & 2 & 2 & 0 & 0 & 2 & 1 & 1 &  &  &  \\
    \hline
    ${M=5}$ (226) & 2 & 2 & 0 & 0 & 2 & 2 & 2 &  &  &  \\
    \hline
    ${M=5}$ (227) & 2 & 2 & 0 & 1 & 2 & 0 & 0 &  &  &  \\
    \hline
    ${M=5}$ (228) & 2 & 2 & 0 & 1 & 2 & 1 & 1 &  &  &  \\
    \hline
    ${M=5}$ (229) & 2 & 2 & 0 & 1 & 2 & 2 & 2 &  &  &  \\
    \hline
    ${M=5}$ (230) & 2 & 2 & 1 & 0 & 1 & 0 & 1 &  &  &  \\
    \hline
    ${M=5}$ (231) & 2 & 2 & 1 & 0 & 1 & 1 & 0 &  &  &  \\
    \hline
    ${M=5}$ (232) & 2 & 2 & 1 & 0 & 1 & 1 & 2 &  &  &  \\
    \hline
    ${M=5}$ (233) & 2 & 2 & 1 & 0 & 1 & 2 & 1 &  &  &  \\
    \hline
    ${M=5}$ (234) & 2 & 2 & 1 & 0 & 1 & 2 & 2 &  &  &  \\
    \hline
    ${M=5}$ (235) & 2 & 2 & 1 & 0 & 2 & 0 & 1 &  &  &  \\
    \hline
    ${M=5}$ (236) & 2 & 2 & 1 & 0 & 2 & 1 & 0 &  &  &  \\
    \hline
    ${M=5}$ (237) & 2 & 2 & 1 & 0 & 2 & 1 & 2 &  &  &  \\
    \hline
    ${M=5}$ (238) & 2 & 2 & 1 & 0 & 2 & 2 & 1 &  &  &  \\
    \hline
    ${M=5}$ (239) & 2 & 2 & 1 & 0 & 2 & 2 & 2 &  &  &  \\
    \hline
    ${M=5}$ (240) & 2 & 2 & 1 & 1 & 1 & 0 & 1 &  &  &  \\
    \hline
    ${M=5}$ (241) & 2 & 2 & 1 & 1 & 1 & 1 & 0 &  &  &  \\
    \hline
    ${M=5}$ (242) & 2 & 2 & 1 & 1 & 1 & 1 & 2 &  & \checkmark &  \\
    \hline
    ${M=5}$ (243) & 2 & 2 & 1 & 1 & 1 & 2 & 1 &  &  &  \\
    \hline
    ${M=5}$ (244) & 2 & 2 & 1 & 1 & 1 & 2 & 2 &  &  &  \\
    \hline
    ${M=5}$ (245) & 2 & 2 & 1 & 1 & 2 & 0 & 1 &  &  &  \\
    \hline
    ${M=5}$ (246) & 2 & 2 & 1 & 1 & 2 & 1 & 0 &  &  &  \\
    \hline
    ${M=5}$ (247) & 2 & 2 & 1 & 1 & 2 & 1 & 2 &  &  &  \\
    \hline
    ${M=5}$ (248) & 2 & 2 & 1 & 1 & 2 & 2 & 1 &  &  &  \\
    \hline
    ${M=5}$ (249) & 2 & 2 & 1 & 1 & 2 & 2 & 2 &  &  &  \\
    \hline
    ${M=5}$ (250) & 2 & 2 & 2 & 0 & 0 & 0 & 2 &  & \checkmark &  \\
    \hline
    ${M=5}$ (251) & 2 & 2 & 2 & 0 & 0 & 1 & 1 &  &  &  \\
    \hline
    ${M=5}$ (252) & 2 & 2 & 2 & 0 & 0 & 1 & 2 &  &  &  \\
    \hline
    ${M=5}$ (253) & 2 & 2 & 2 & 0 & 0 & 2 & 0 &  &  &  \\
    \hline
    ${M=5}$ (254) & 2 & 2 & 2 & 0 & 0 & 2 & 1 &  &  &  \\
    \hline
    ${M=5}$ (255) & 2 & 2 & 2 & 0 & 1 & 0 & 2 &  &  &  \\
    \hline
    ${M=5}$ (256) & 2 & 2 & 2 & 0 & 1 & 1 & 1 &  &  &  \\
    \hline
    ${M=5}$ (257) & 2 & 2 & 2 & 0 & 1 & 1 & 2 &  &  &  \\
    \hline
    ${M=5}$ (258) & 2 & 2 & 2 & 0 & 1 & 2 & 0 &  &  &  \\
    \hline
    ${M=5}$ (259) & 2 & 2 & 2 & 0 & 1 & 2 & 1 &  &  &  \\
    \hline
    ${M=5}$ (260) & 2 & 2 & 2 & 1 & 0 & 0 & 2 &  &  &  \\
    \hline
    ${M=5}$ (261) & 2 & 2 & 2 & 1 & 0 & 1 & 1 &  &  &  \\
    \hline
    ${M=5}$ (262) & 2 & 2 & 2 & 1 & 0 & 1 & 2 &  &  &  \\
    \hline
    ${M=5}$ (263) & 2 & 2 & 2 & 1 & 0 & 2 & 0 &  &  &  \\
    \hline
    ${M=5}$ (264) & 2 & 2 & 2 & 1 & 0 & 2 & 1 &  &  &  \\
    \hline
    ${M=5}$ (265) & 2 & 2 & 2 & 1 & 1 & 0 & 2 &  &  &  \\
    \hline
    ${M=5}$ (266) & 2 & 2 & 2 & 1 & 1 & 1 & 1 &  &  &  \\
    \hline
    ${M=5}$ (267) & 2 & 2 & 2 & 1 & 1 & 1 & 2 &  & \checkmark &  \\
    \hline
    ${M=5}$ (268) & 2 & 2 & 2 & 1 & 1 & 2 & 0 &  &  &  \\
    \hline
    ${M=5}$ (269) & 2 & 2 & 2 & 1 & 1 & 2 & 1 &  &  &  \\
    \hline
\end{longtable}

\newpage

\begin{tiny}
\begin{longtable}{|p{1.5cm} |c |c |c |c |c ||c |c |c |c |c|}
    \caption{Allowed operators in F-terms for $M=5$, represented by the symbol $\checkmark$ for each case in Table~\ref{tab:assign-M=5}.}
    \label{tab:allowed-op-M=5}  \\
        
    \hline
    Case & $H_dH_u$ & $LH_u$ & $LL\bar{E}$ & $LQ\bar{D}$ & $\bar{U}\bar{D}\bar{D}$ & $QQQL$ & $\bar{U}\bar{U}\bar{D}\bar{E}$ & $QQQH_d$ & $Q\bar{U}\bar{E}H_d$ & $LH_uH_dH_u$ \\
    \hline\hline
    \endfirsthead
    
    \hline
    Case & $H_dH_u$ & $LH_u$ & $LL\bar{E}$ & $LQ\bar{D}$ & $\bar{U}\bar{D}\bar{D}$ & $QQQL$ & $\bar{U}\bar{U}\bar{D}\bar{E}$ & $QQQH_d$ & $Q\bar{U}\bar{E}H_d$ & $LH_uH_dH_u$ \\
    \hline\hline
    \endhead
    
    (1), (69), (77), (140), (142), (145), (147), (189), (192), (194), (197), (250), (257) & \checkmark & \checkmark & \checkmark & \checkmark & \checkmark & \checkmark & \checkmark & \checkmark & \checkmark &  \checkmark \\
    \hline
    (2), (3), (70), (75), (253), (258) & \checkmark &  &  &  & \checkmark &  &  & \checkmark &  &  \\
    \hline
    (4), (7), (9), (11), (63), (224) &  &  & \checkmark & \checkmark & \checkmark &  &  &  &  & \checkmark \\
    \hline
    (5), (12), (64), (67), (226), (229) &  & \checkmark &  &  & \checkmark & \checkmark & \checkmark &  & \checkmark &   \\
    \hline
    (6), (13) &  &  & \checkmark &  & \checkmark &  & \checkmark &  & \checkmark &   \\
    \hline
    (8), (10) &  &  &  &  & \checkmark &  &  &  &  &  \\
    \hline
    (14), (27), (43), (115), (168), (214) &  &  & \checkmark & \checkmark &  & \checkmark & \checkmark & \checkmark &  & \checkmark \\
    \hline
    (15), (29), (44), (171) &  & \checkmark &  &  &  &  &  & \checkmark & \checkmark &  \\
    \hline
    (16), (28) &  &  &  &  &  &  & \checkmark & \checkmark &  & \checkmark \\
    \hline
    \rowcolor{blue!20}
    (17), (35), (40), (155) & \checkmark & \checkmark & \checkmark & \checkmark &  &  &  &  & \checkmark & \checkmark \\
    \hline
    \rowcolor{red!20}
    (18), (38), (41), (157) & \checkmark &  &  &  &  & \checkmark & \checkmark &  &  &  \\
    \hline
    (19), (39) & \checkmark &  & \checkmark &  &  &  & \checkmark &  &  & \checkmark \\
    \hline
    (20), (37) & \checkmark & \checkmark & \checkmark & \checkmark &  &  & \checkmark &  & \checkmark  & \checkmark \\
    \hline
    (21), (36) & \checkmark &  &  &  &  &  & \checkmark &  & \checkmark & \checkmark \\
    \hline
    (22), (30), (66), (112), (178), (227) &  &  & \checkmark & \checkmark & \checkmark &  & \checkmark &  &  & \checkmark \\
    \hline
    (23), (34) &  & \checkmark &  &  & \checkmark &  & \checkmark &  & \checkmark & \checkmark \\
    \hline
    (24), (33) &  &  & \checkmark & \checkmark & \checkmark &  & \checkmark &  & \checkmark & \checkmark \\
    \hline
    (25), (31), (113), (180) &  &  &  &  & \checkmark & \checkmark & \checkmark &  & \checkmark & \checkmark \\
    \hline
    (26), (32) &  & \checkmark & \checkmark &  & \checkmark &  & \checkmark &  & \checkmark &\checkmark \\
    \hline
    (42), (156) & \checkmark &  &  & \checkmark &  & \checkmark & \checkmark &  & \checkmark &  \\
    \hline
    (45), (172) &  &  & \checkmark &  &  & \checkmark &  & \checkmark & \checkmark &  \\
    \hline
    (46), (55), (106), (118), (125), (133), (166), (170), (204), (211), (216), (243) &  &  & \checkmark & \checkmark &  & \checkmark & \checkmark & \checkmark & \checkmark &  \checkmark \\
    \hline
    (47), (54), (167), (169) &  &  &  &  &  & \checkmark & \checkmark & \checkmark & \checkmark &  \\
    \hline
    (48), (120), (173), (219) &  &  & \checkmark & \checkmark & \checkmark & \checkmark & \checkmark & \checkmark &  & \checkmark \\
    \hline
    (49), (61), (159), (176) &  & \checkmark &  &  & \checkmark &  & \checkmark & \checkmark & \checkmark &  \\
    \hline
    (50), (62), (160), (177) &  &  & \checkmark & \checkmark & \checkmark & \checkmark & \checkmark & \checkmark & \checkmark &   \\
    \hline
    (51), (58), (60), (91), (99), (101), (123), (158), (161), (175), (221), (238), (245), (248) &  &  & \checkmark & \checkmark & \checkmark & \checkmark & \checkmark & \checkmark & \checkmark &\checkmark \\
    \hline
    (52), (59), (162), (174) &  &  &  & \checkmark & \checkmark & \checkmark & \checkmark & \checkmark & \checkmark &  \\
    \hline
    (53), (104), (128), (130), (163), (206), (209), (240) &  &  & \checkmark & \checkmark &  & \checkmark &  & \checkmark & \checkmark &\checkmark \\
    \hline
    (56), (164) &  & \checkmark &  & \checkmark &  &  & \checkmark & \checkmark & \checkmark &  \\
    \hline
    (57), (165) &  &  & \checkmark &  &  & \checkmark & \checkmark & \checkmark & \checkmark &  \\
    \hline
    (65), (225) &  &  &  & \checkmark & \checkmark & \checkmark &  &  &  & \checkmark \\
    \hline
    (68), (228) &  &  &  & \checkmark & \checkmark & \checkmark & \checkmark &  & \checkmark & \checkmark \\
    \hline
    (71), (254) & \checkmark &  & \checkmark &  & \checkmark & \checkmark &  & \checkmark &  &\checkmark \\
    \hline
    (72), (74), (252), (255) & \checkmark & \checkmark & \checkmark & \checkmark & \checkmark & \checkmark &  & \checkmark & \checkmark &  \checkmark \\
    \hline
    (73), (251) & \checkmark &  &  &  & \checkmark & \checkmark &  & \checkmark & \checkmark &  \checkmark \\
    \hline
    (76), (259) & \checkmark &  & \checkmark & \checkmark & \checkmark & \checkmark &  & \checkmark &  & \checkmark \\
    \hline
    (78), (141), (198), (256) & \checkmark &  &  & \checkmark & \checkmark & \checkmark & \checkmark & \checkmark & \checkmark & \checkmark \\
    \hline
    (79), (87), (137), (152), (187), (202), (260), (267) & \checkmark & \checkmark & \checkmark & \checkmark &  & \checkmark & \checkmark & \checkmark & \checkmark & \checkmark \\
    \hline
    (80), (85), (153), (185), (263), (268) & \checkmark &  &  &  &  &  & \checkmark & \checkmark & \checkmark &  \\
    \hline
    (81), (139), (201), (264) & \checkmark &  & \checkmark &  &  & \checkmark & \checkmark & \checkmark & \checkmark & \checkmark \\
    \hline
    (82), (84), (135), (150), (184), (199), (262), (265) & \checkmark & \checkmark & \checkmark & \checkmark &  & \checkmark &  & \checkmark & \checkmark & \checkmark \\
    \hline
    (83), (261) & \checkmark &  &  &  &  & \checkmark &  & \checkmark & \checkmark & \checkmark \\
    \hline
    (86), (154), (186), (269) & \checkmark &  & \checkmark & \checkmark &  & \checkmark & \checkmark & \checkmark & \checkmark & \checkmark \\
    \hline
    (88), (151), (188), (266) & \checkmark &  &  & \checkmark &  & \checkmark & \checkmark & \checkmark & \checkmark &  \checkmark \\
    \hline
    (89), (94), (230), (235) &  &  & \checkmark & \checkmark & \checkmark & \checkmark &  & \checkmark &  & \checkmark \\
    \hline
    (90), (100), (239), (249) &  & \checkmark &  &  & \checkmark & \checkmark & \checkmark & \checkmark & \checkmark &  \checkmark \\
    \hline
    (92), (236) &  &  &  & \checkmark & \checkmark &  &  & \checkmark & \checkmark & \checkmark \\
    \hline
    (93), (237) &  & \checkmark & \checkmark &  & \checkmark & \checkmark &  & \checkmark & \checkmark & \checkmark \\
    \hline
    (95), (234) &  & \checkmark &  & \checkmark & \checkmark & \checkmark &  & \checkmark & \checkmark & \checkmark \\
    \hline
    (96), (233) &  &  & \checkmark & \checkmark & \checkmark & \checkmark &  & \checkmark & \checkmark & \checkmark \\
    \hline
    (97), (121), (222), (231) &  &  &  &  & \checkmark &  & \checkmark & \checkmark & \checkmark &  \checkmark \\
    \hline
    (98), (122), (223), (232) &  & \checkmark & \checkmark & \checkmark & \checkmark & \checkmark & \checkmark & \checkmark & \checkmark & \checkmark \\
    \hline
    (102), (246) &  &  &  & \checkmark & \checkmark &  & \checkmark & \checkmark & \checkmark & \checkmark \\
    \hline
    (103), (247) &  & \checkmark & \checkmark &  & \checkmark & \checkmark & \checkmark & \checkmark &  \checkmark & \checkmark \\
    \hline
    (105), (134), (210), (244) &  & \checkmark &  & \checkmark &  & \checkmark & \checkmark & \checkmark & \checkmark & \checkmark \\
    \hline
    (107), (126), (131), (207), (212), (241) &  &  &  &  &  &  & \checkmark & \checkmark & \checkmark & \checkmark \\
    \hline
    (108), (132), (213), (242) &  & \checkmark & \checkmark & \checkmark &  & \checkmark & \checkmark & \checkmark & \checkmark & \checkmark \\
    \hline
    (109), (181) &  &  & \checkmark & \checkmark &  &  &  &  & \checkmark & \checkmark \\
    \hline
    (110), (183) &  &  &  &  &  & \checkmark & \checkmark &  &  & \checkmark \\
    \hline
    (111), (182) &  & \checkmark &  & \checkmark &  & \checkmark & \checkmark &  & \checkmark &  \\
    \hline
    (114), (179) &  & \checkmark &  & \checkmark & \checkmark & \checkmark & \checkmark &  & \checkmark &  \\
    \hline
    (116), (217) &  &  &  &  &  &  &  & \checkmark & \checkmark & \checkmark \\
    \hline
    (117), (218) &  & \checkmark & \checkmark &  &  & \checkmark &  & \checkmark & \checkmark & \checkmark \\
    \hline
    (119), (215) &  & \checkmark &  &  &  & \checkmark & \checkmark & \checkmark & \checkmark & \checkmark \\
    \hline
    (124), (220) &  & \checkmark &  & \checkmark & \checkmark & \checkmark & \checkmark & \checkmark & \checkmark & \checkmark \\
    \hline
    (127), (208) &  & \checkmark & \checkmark &  &  & \checkmark & \checkmark & \checkmark & \checkmark & \checkmark \\
    \hline
    (129), (205) &  & \checkmark &  &  &  & \checkmark &  & \checkmark & \checkmark & \checkmark \\
    \hline
    (136), (203) & \checkmark &  &  &  &  & \checkmark & \checkmark & \checkmark & \checkmark & \checkmark \\
    \hline
    (138), (200) & \checkmark &  &  & \checkmark &  &  & \checkmark & \checkmark & \checkmark &   \\
    \hline
    (143), (195) & \checkmark &  &  &  & \checkmark &  & \checkmark & \checkmark & \checkmark &   \\
    \hline
    (144), (196) & \checkmark &  & \checkmark & \checkmark & \checkmark & \checkmark & \checkmark & \checkmark & \checkmark & \checkmark \\
    \hline
    (146), (193) & \checkmark &  &  &  & \checkmark & \checkmark & \checkmark & \checkmark & \checkmark & \checkmark \\
    \hline
    (148), (190) & \checkmark &  &  & \checkmark & \checkmark &  & \checkmark & \checkmark & \checkmark &  \\
    \hline
    (149), (191) & \checkmark &  & \checkmark &  & \checkmark & \checkmark & \checkmark & \checkmark & \checkmark & \checkmark \\
    \hline
\end{longtable}    
\end{tiny}

\begin{tiny}
    \begin{longtable}{|p{2.5cm} |c |c |c |c|}
        \caption{Allowed operators in D-terms represented by the symbol $\checkmark$ for $M=5$.}
        \label{tab:allowed-op-M=5D}  \\
        \hline
        Case & $\bar{U}\bar{D}^*\bar{E}$ & $H_u^*H_d\bar{E}$ & $Q\bar{U}L^*$ & $QQ\bar{D}^*$ \\\hline
        \endfirsthead
        \hline
        Case & $\bar{U}\bar{D}^*\bar{E}$ & $H_u^*H_d\bar{E}$ & $Q\bar{U}L^*$ & $QQ\bar{D}^*$ \\\hline
        \endhead
        \hline
        \endfoot
        \hline
        \endlastfoot
        (1), (15), (29), (44), (45), (49), (50), (61), (69), (77), (82), (83), (84), (98), (105), (106), (116), (117), (121), (122), (124), (128), (129), (132), (142), (150), (151), (152), (159), (171), (172), (176), (177), (184), (187), (188), (197), (205), (206), (213), (217), (218), (220), (222), (223), (232), (243), (244), (250), (257), (261), (262), (265) & \checkmark & \checkmark & \checkmark & \checkmark \\\hline
        (2), (3), (14), (16), (27), (28), (43), (46), (48), (70), (71), (75), (76), (80), (94), (115), (120), (153), (168), (170), (173), (185), (214), (219), (230), (253), (254), (258), (259), (263) &  &  &  & \checkmark \\\hline
        (4), (7), (8), (9), (10), (11), \colorbox{red!20}{(18)}, (19), (22), (30), \colorbox{red!20}{(38)}, (39), \colorbox{red!20}{(41)}, (54), (63), (65), (66), (89), (110), (112), \colorbox{red!20}{(157)}, (167), (178), (183), (224), (225), (227), (235) &  &  &  &  \\\hline
        (5), (12), \colorbox{blue!20}{(17)}, (20), (23), (26), (32), (34), \colorbox{blue!20}{(35)}, (37), \colorbox{blue!20}{(40)}, (56), (64), (67), (90), (100), (101), (103), (111), (114), (135), (146), (147), \colorbox{blue!20}{(155)}, (164), (179), (182), (192), (193), (199), (226), (229), (239), (247), (248), (249) & \checkmark & \checkmark & \checkmark &  \\\hline
        (6), (13), (21), (24), (25), (31), (33), (36), (57), (91), (102), (165), (238), (246) & \checkmark & \checkmark &  &  \\\hline
        (42), (53), (68), (109), (113), (138), (149), (156), (163), (180), (181), (191), (200), (228) & \checkmark &  & \checkmark &  \\\hline
        (47), (58), (81), (125), (154), (158), (169), (186), (204), (264) &  &  & \checkmark & \checkmark \\\hline
        (51), (59), (85), (144), (162), (175), (196), (268) & \checkmark &  &  & \checkmark \\\hline
        (52), (60), (86), (104), (130), (143), (161), (174), (195), (209), (240), (269) & \checkmark &  & \checkmark & \checkmark \\\hline
        (55), (99), (166), (245) &  &  & \checkmark &  \\\hline
        (62), (78), (97), (123), (131), (141), (160), (198), (212), (221), (231), (256) & \checkmark & \checkmark &  & \checkmark \\\hline
        (72), (74), (79), (87), (88), (95), (108), (119), (127), (133), (134), (140), (194), (208), (210), (211), (215), (234), (242), (252), (255), (260), (266), (267) &  & \checkmark & \checkmark & \checkmark \\\hline
        (73), (96), (107), (118), (126), (207), (216), (233), (241), (251) &  & \checkmark &  & \checkmark \\\hline
        (92), (136), (203), (236) &  & \checkmark &  &  \\\hline
        (93), (137), (145), (189), (202), (237) &  & \checkmark & \checkmark &  \\\hline
        (139), (148), (190), (201) & \checkmark &  &  &  \\\hline
    \end{longtable}
\end{tiny}

\bibliography{references}{}
\bibliographystyle{JHEP}

\end{document}